\documentclass[final,5p,twocolumn,sort&compress]{elsarticle}

\usepackage{amssymb}
\usepackage{color}
\usepackage[switch]{lineno}
\makeatletter
\def\ps@pprintTitle{%
 \let\@oddhead\@empty
 \let\@evenhead\@empty
 \def\@oddfoot{}%
 \let\@evenfoot\@oddfoot}

\journal{Journal of Colloid and Interface Science}

\begin{document}

\begin{frontmatter}

\title{Thermoresponsivity of poly(N-isopropylacrylamide) microgels in water-trehalose solution and its relation to protein behavior}


\author[label1]{Benedetta Petra Rosi\fnref{myfootnote}}
\address[label1]{Dipartimento di Fisica e Geologia, Universit\`{a} di Perugia, I-06123 Perugia, Italy}
\fntext[myfootnote]{These authors contributed equally}

\author[label2,label3]{Letizia Tavagnacco\fnref{myfootnote}}
\address[label2]{CNR-ISC, Sapienza Universit\`{a} di Roma,  I-00185 Roma, Italy}
\address[label3]{Dipartimento di Fisica, Sapienza Universit\`{a} di Roma, I-00185 Roma, Italy}

\author[label4]{Lucia Comez}
\address[label4]{CNR-IOM, Dipartimento di Fisica e Geologia, Universit\`{a} di Perugia, I-06123 Perugia, Italy}

\author[label5]{Paola Sassi}
\address[label5]{Dipartimento di Chimica, Biologia e Biotecnologie, Universit\`{a} di Perugia, I-06123 Perugia, Italy}

\author[label5]{Maria Ricci}

\author[label2,label3]{Elena Buratti}

\author[label7,label8]{Monica Bertoldo}
\address[label7]{Dipartimento di Scienze Chimiche, Farmaceutiche ed Agrarie, Universit\`{a} di Ferrara, I-44121 Ferrara, Italy}
\address[label8]{CNR-ISOF, Area della Ricerca, I-40129 Bologna, Italy}

\author[label1]{Caterina Petrillo}

\author[label2,label3]{Emanuela Zaccarelli}

\author[label9]{Ester Chiessi\corref{mycorrespondingauthor}}
\address[label9]{Dipartimento di Scienze e Tecnologie Chimiche, Universit\`{a} di Roma ``Tor Vergata'', I-00133 Roma, Italy}
\ead{ester.chiessi@uniroma2.it}

\author[label1]{Silvia Corezzi\corref{mycorrespondingauthor}}
\cortext[mycorrespondingauthor]{Corresponding author}
\ead{silvia.corezzi@unipg.it}

\begin{abstract}

\textit{Hypotheses:} Additives are commonly used to tune
macromolecular conformational transitions. Among additives,
trehalose is an excellent bioprotectant and among responsive
polymers, PNIPAM is the most studied material. Nevertheless, their
interaction mechanism so far has only been hinted without direct
investigation, and, crucially, never elucidated in comparison to
proteins. Detailed insights would help understand to what extent
PNIPAM microgels can effectively be used as synthetic biomimetic
materials, to reproduce and study, at the colloidal scale, isolated
protein behavior and its sensitivity to interactions with specific
cosolvents or cosolutes. \textit{Experiments:} The effect of
trehalose on the swelling behavior of PNIPAM microgels was monitored
by dynamic light scattering; Raman spectroscopy and molecular
dynamics simulations were used to explore changes of solvation and
dynamics across the swelling-deswelling transition at the molecular
scale. \textit{Findings:} Strongly hydrated trehalose molecules
develop water-mediated interactions with PNIPAM microgels, thereby
preserving polymer hydration below and above the transition while
drastically inhibiting local motions of the polymer and of its
hydration shell. Our study, for the first time, demonstrates that
slowdown of dynamics and preferential exclusion are the principal
mechanisms governing trehalose effect on PNIPAM microgels, at odds
with preferential adsorption of alcohols, but in full analogy with
the behavior observed in trehalose-protein systems.

\end{abstract}

\begin{keyword}
Poly(N-isopropylacrylamide) (PNIPAM) \sep microgels \sep trehalose
\sep cosolvents and cosolutes \sep lower critical solution
temperature (LCST) \sep volume phase transition \sep biomimetic
material \sep bioprotection \sep hydration water


\end{keyword}

\end{frontmatter}

\section{\label{Intro}Introduction}

The conformation of synthetic and biological macromolecules is known
to exhibit a remarkably strong dependence on temperature and other
environmental conditions, such as the presence of additives in the
solvent. The example par excellence in biology is the denaturation
of proteins, for which even a minor change of temperature may result
in an abrupt change of structure and loss of biological activity. In
globular proteins the destruction of the native functional
conformation sometimes gives rise to a fully unfolded (random coil)
state, but in a number of cases occurs through partially unfolded
states, which are compact and preserve the general architecture of
the native structure \cite{BookProteinPhysics}. It is well-known
that such conformational transitions of proteins in solution are
strongly affected by the interaction with cosolvents or cosolutes,
which perform, and are able to tune, biologically relevant functions
such as cryoconservation, bioprotection, stabilization,
denaturation, to name a few \cite{BookProteinInteractions}. On the
other hand, temperature-induced coil-globule transitions also occur
in synthetic macromolecules, the simplest case being that of
homopolymers like poly($N$-isopropylacrylamide) (PNIPAM). This
polymer, dispersed in water, undergoes a reversible transition at a
lower critical solution temperature (LCST, $\sim$305 K) from a coil
conformation of the chains for $T<$LCST to a globule conformation
above it. Such intramolecular collapse is also highly sensitive to
the presence of additives, with the LCST being shifted by organic
molecules \cite{CostaPOLYMER2002, YamauchiJPCB2007, PerezAPM2019},
salts \cite{ZhangJACS2005, HeydaJPCB2014, PicaPCCP2020}, osmolytes
\cite{ShpigelmanJPSB2008, SagleJACS2009, KimEPJ2014,
SchroerPCCP2016, NarangNJC2018}, ionic liquids
\cite{UmapathiJML2020}, and proteins \cite{UmapathiJCIS2019}.
Additives affect the thermoresponsivity of PNIPAM-based copolymers
as well \cite{NarangACIS2019, MartinezMoroJCIS2020, KumarLANG2021}.
Moreover, some experiments have shown that PNIPAM displays
additional `protein-like' behavioral traits, such as a cooperative
character of the coil-globule transition \cite{TiktopuloMACRO1994}
and the occurrence of a dynamical transition at low temperature
\cite{ZanattaSCIADV2018}. Altogether, these features make PNIPAM
potentially very useful as a synthetic biomimetic material, that
could be employed as a simplified model to elucidate phenomena of
high biological relevance. In this context, it would be crucial to
shed light on the extent to which PNIPAM is capable of mimicking the
protein sensitivity to the interaction with specific cosolvents or
cosolutes.

To this aim, in the present study we focus on trehalose, an
excellent bioprotectant, and use PNIPAM microgels to mimic the
complex macromolecular environment of globular proteins, determining
whether the additive-induced effects on PNIPAM microgels and
proteins in solution occur via similar molecular mechanisms.
Microgels are cross-linked polymer networks of colloidal size, in
which the temperature-induced coil-globule transition of the
constituent polymer chains reflects into a drastic volume change,
i.e. an ability to swell and deswell (so-called volume phase
transition) \cite{BookFernandez2011}. Owing to the combination of
being responsive while still having a stable crosslinked
architecture these particles are, on one hand, particularly suited
to represent a globular protein during the unfolding process and, on
the other hand, able to amplify the effect of trehalose on
PNIPAM-water interactions on a colloidal scale, by making it visible
through the effect on the microgel hydrodynamic size.

Trehalose is a natural disaccharide, whose outstanding bioprotective
capabilities have long been recognized and increasingly used as a
way of preserving the biological activity of proteins and enzymes
under physicochemical stress conditions that would normally promote
lability or denaturation, such as cryogenic temperatures
\cite{BeattieDIAB1997}, heating \cite{CarninciPNAS1998}, presence of
chemicals \cite{BenaroudjJBC2001}, or dehydration
\cite{CesaroNATMAT2006}. Several studies contributed over the years
to shed light on the molecular mechanisms governing the trehalose
effectiveness in preserving the functional conformation of proteins
in solution \cite{KaushikJBC2003, JainPROTSCI2009}, making it clear
that the extent of preferential \textit{exclusion} of the sugar from
the protein hydration shell, linked to protein preferential
hydration \cite{TimasheffBIOCHEM1982, TimasheffBIOCHEM1997,
KaushikJBC2003}, and the \textit{slowdown} caused to local motions
of the protein and to its hydration water are the major contributing
factors \cite{CiceroneBIOPHYSJ2004, SubatraJPCB2015,
MalferrariJPCL2016}. Recent studies on lysozyme
\cite{CorradiniSCIREP2013, CorezziJCP2019, CamisascaJCP2020} have
shown that these mechanisms may be associated, upon cooling, to a
remarkably enhanced slowdown of water molecules located in proximity
to the protein surface, an effect that inhibits ice formation and
facilitates vitrification without biological damage, therefore
providing clues as to how a cryoprotectant action of trehalose is
also possible.

Aqueous PNIPAM ternary systems have so far been studied by means of
various techniques and for a number of additives, yet this work
represents the first direct investigation of the PNIPAM-trehalose
interaction mechanism. Previous studies in water-trehalose mixture
are scarce and only concerned linear PNIPAM chains, where a decrease
of LCST, and hence the promotion of the globular compact state of
the polymer, on increasing trehalose molar concentration was found
to be more pronounced than in presence of other sugars
\cite{ShpigelmanJPSB2008, NarangJCIS2017}. In this work we monitor
the volume phase transition of PNIPAM microgels comparatively in
water and in water-trehalose mixture at 0.72 M sugar concentration,
a value that is able to significantly lower the polymer LCST
compared to that in water, and that falls well within the range of
concentrations used to investigate the trehalose stabilization
effect of proteins in solution \cite{TimasheffBIOCHEM1997,
KaushikJBC2003}. Microgels are synthesized with low crosslinking
degree, to have a high swelling capacity \cite{Karg2013}, and with a
small size (less than 50 nm in hydrodynamic radius at room
temperature \cite{ArlethJPSB2005, AnderssonJPSB2006}), to minimize
heterogeneity of the internal structure. By combining dynamic light
scattering (DLS) and Raman spectroscopy (RS) experiments with
all-atom molecular dynamics (MD) simulations, we correlate the
conformational transition of the particles detected at the colloidal
length scale with changes of structural and dynamical properties,
probed at the atomic scale. In particular RS measurements, sensitive
to the solvation pattern and intramolecular bonding of microgels,
are complemented by simulations which are able to elucidate the
influence of trehalose on the water affinity for PNIPAM, the
specific and selective interactions, if any, of the sugar with the
polymer and, very importantly, how the presence of trehalose affects
the dynamics of the system, including polymer, bulk and hydration
water. We find that strongly hydrated trehalose molecules mainly
develop water-mediated interactions with PNIPAM microgels, making
trehalose preferentially excluded from the polymer surface,
especially in the swollen state. Preferential exclusion coexists
with the binding of some trehalose to the polymer, which occurs,
however, without interfering with the hydration pattern of the
microgel network. Hydration is preserved both before and after the
transition, in a way that the amount of absorbed water remains the
same as without trehalose, and the composition of the absorbed
solvent remains unaffected by the conformational transition of the
particles. Meanwhile, sugar molecules also stabilize microgels by
drastically inhibiting local motions of the polymer and of its
hydration shell, as well as by increasing the residence time of
water in the polymer surrounding. Overall, the present results
highlight that trehalose promotes the collapsed conformation of
PNIPAM microgels similarly to the way it stabilizes aqueous proteins
in their native globular state. It also induces molecular mechanisms
that are similar to those governing its own action against protein
damage during freezing. This confirms and extends the use of PNIPAM
as a suitable biomimetic material of a synthetic nature.

\section{\label{Methods}Materials and Methods}

      \subsection{Materials}

The monomer $N$-isopropylacrylamide (NIPAM) ($M_w$=113.16,
Sigma-Aldrich, 97 $\%$ purity) and the crosslinker
$N$,$N$'-methylene-bis-acrylamide (BIS) ($M_w$=154.17, Eastman
Kodak, electrophoresis grade) were purified by recrystallization
from hexane and methanol, respectively, then dried under reduced
pressure (0.01 mmHg) at room temperature and stored at 253 K until
use. The surfactant sodium dodecyl sulphate (SDS) ($M_w$=288.372,
98$\%$ purity), the initiator potassium persulfate (KPS)
($M_w$=270.322, 98$\%$ purity) and all solvents (RP grade) were
purchased from Sigma-Aldrich and used as received. Ultrapure water
(resistivity: 18.2 M$\Omega$/cm at 298 K) was obtained with a
Millipore Direct-Q\textsuperscript{\textregistered} 3 UV
purification system. Dialysis tubing cellulose membrane
(Sigma-Aldrich), 14 kDa molecular weight cut-off, was cleaned before
use in running distilled water for 3 hours, treated in a solution of
sodium hydrogen carbonate (NaHCO$_3$, 3.0 $\%$ wt) and
ethylenediaminetetraacetic acid (EDTA, 0.4 $\%$ wt) at 343 K for 10
min, rinsed in distilled water at 343 K for 10 min and finally in
fresh distilled water at room temperature for 2 h.

      \subsection{Synthesis of PNIPAM microgels}

Microgels were synthesized by precipitation polymerization in a 2000
mL four-necked jacked reactor equipped with condenser and mechanical
stirrer. Proper amounts of NIPAM (24.16(2) g), BIS (0.448(0) g) and
SDS (3.519(0) g) were dissolved in 1560 mL of ultrapure water and
transferred into the reactor, where the solution was deoxygenated by
bubbling nitrogen for 1 h and then heated at $343\pm 1$ K. KPS
(1.037(6) g, dissolved in 20 mL of deoxygenated water) was added to
initiate the polymerization, and the reaction was carried out for 16
h. The resultant PNIPAM microgel particles were purified by dialysis
against distilled water with frequent water change for 2 weeks, and
then lyophilized to constant weight.

Since the SDS concentration (7.72 mM) was slightly below the
critical micelle concentration (8.18 mM), small nanosized microgels
were obtained \cite{BookFernandez2011}. In contrast to bigger
microgels, characterized by a heterogeneous structure with a dense
core and a loose corona, the small-sized microgels have an almost
uniform radial crosslinker density \cite{ArlethJPSB2005,
AnderssonJPSB2006}. The average chain length between crosslinking
points is about 37 PNIPAM residues. As the size of a trehalose
molecule is comparable to that of three/four PNIPAM repeating units
(see sec. \label{MD} below), the mesh size of the microgel network
does not hinder the disaccharide to permeate the particles.

       \subsection{Preparation of microgel suspensions}

Lyophilized PNIPAM microgels were re-suspended in ultrapure water
and in water-trehalose solution. D-(+)-trehalose dihydrate
($M_w$=378.33, Sigma-Aldrich, $\geq$ 99.0 $\%$ purity) was used as
received. The water-trehalose solution was prepared by weighing
(24.9 wt$\%$ of trehalose dihydrate). The resulting concentration of
trehalose was 0.72 M. In terms of number of molecules, the trehalose
to water mole ratio was $1:65.3$, or equivalently, the trehalose
mole fraction was $x_{tr}=0.015$. The solution was heated at 313 K
and continuously stirred until complete trehalose dissolution, then
thermalized at room temperature and filtered through a cellulose
membrane filter, 0.2 $\mu$m pore size.

For DLS experiments, microgel suspensions in the two solvents were
prepared at high dilution (microgel concentration $\sim
10^{-4}-10^{-5}$ g/mL). For RS experiments and for thermal analysis,
samples were prepared at microgel concentration of 18 wt$\%$ and 3
wt$\%$, respectively. All samples were stored at 275--277 K for at
least two days before use.

     \subsection{Characterization of solvent media}

Both water and the water-trehalose solution used to prepare microgel
suspensions were characterized by density, viscosity and refractive
index, over the temperature range of DLS measurements.

For the water-trehalose solution, the kinematic viscosity $\nu$ was
measured with a micro-Ubbelhode viscosimeter, the mass density
$\rho$ with an Anton Paar DMA 5000 M densitometer, and the dynamic
viscosity $\eta$ was calculated as $\eta=\rho \nu$. At any
temperature, the refractive index $n_{D}$ is related to $\rho$ by
the Lorentz-Lorenz equation,
$r=\rho^{-1}(n_{D}^{2}-1)/(n_{D}^{2}+2)$, where the specific
refractivity $r$ is rather temperature-independent. The value of $r$
was obtained at 298.3 K from $n_{D}=1.368$, measured with a NAR-1T
Liquid Abbe refractometer, and $\rho=1.098643$ g cm$^{-3}$,
independently measured.

For pure water, mass density, refractive index, and viscosity data
as a function of temperature are available from the literature. We
used $\rho$ and $n_{D}$ taken from ref. \cite{PhilippSM2012}, after
having verified the agreement with the measured value at one
reference temperature. Measurements of $\nu$ were repeated,
providing values of $\eta$ in perfect agreement with the literature
ones \cite{KorsonJPC1968}. The temperature dependence of $\rho$,
$\eta$ and $n_{D}$ for the two solvent media is shown in Fig. S1.

     \subsection{Dynamic light scattering experiments}

DLS measurements were performed with a commercial setup equipped
with a Brookhaven BI-9000AT correlator, using a solid state laser
source of $\lambda$=532 nm. The monochromatic beam was focused on
the sample placed in a cylindrical VAT for index matching and
temperature control. The temperature was regulated within 0.1
$^{\circ}$C by a thermostatic circulator. The scattered light was
collected at an angle $\theta=90^{\circ}$, corresponding to a
scattering wave vector $q=(4\pi n/\lambda)\sin(\theta/2)$, where $n$
($\approx n_{D}$) is the refractive index of the sample at the
incident wavelength. The intensity autocorrelation function,
$G^{(2)}(q,t)\equiv \langle I(q,0)I(q,t) \rangle$, was acquired as a
function of temperature on heating across the volume phase
transition of the microgels (from 288 to 320 K for the sample in
water, from 278 to 320 K for the sample in water-trehalose). At each
temperature, the sample was equilibrated for 10 min, and then
measured in at least three different points. The results are
reported as the mean value $\pm$SD.

          \subsubsection{Analysis of DLS spectra}

The autocorrelation function of the scattered field,
$G^{(1)}(q,t)\equiv \langle E^{*}(q,0)E(q,t)\rangle$, is linked to
the measured $G^{(2)}(q,t)$ by the Siegert relation
$G^{(2)}(t)=A_{0}\left[1 + \beta |G^{(1)}(t)|^{2} \right]$, with
$A_{0}$ a measured baseline and $\beta$ the coherence factor, an
instrumental parameter of the order of unit. Therefore,
$[G^{(2)}(t)-A_{0}]^{1/2}$ is proportional to $G^{(1)}(t)$. This
quantity was analyzed by the method of cumulants (see text in the
SI), to obtain the intensity-weighted (or z-average) particle
diffusion coefficient $<D>_{z}$. The hydrodynamic radius $R_{h}$ of
the microgel particles in suspension was determined as the z-average
particle size by using the Stokes-Einstein relation
$R_{h}=k_{B}T/6\pi\eta <D>_{z}$, where $k_{B}$ is the Boltzmann's
constant, $T$ the absolute temperature, and $\eta$ the viscosity of
the dispersing medium.

The mentioned analysis, however, only applies if the solvent
molecules are light scatterers far less efficient than the suspended
particles, and if particles move much more slowly than solvent
molecules so that the solvent contribution to the decay of the
correlation function occurs on a time scale smaller than the
experimental time window. It must be noted that these conditions are
fulfilled for the microgel suspension in water, but it is not the
case for the sample in water-trehalose solution. In this latter, the
relaxation dynamics of trehalose molecules is not sufficiently
separated from the motion of microgels, thus entering the time
window of the autocorrelation function and contributing with an
additional decay process to $G^{(1)}(t)$ at small times (see Fig.
S2). The solvent contribution must be carefully subtracted from the
acquired scattering signal, before applying the cumulant method. The
analysis of DLS spectra in the presence of trehalose is illustrated
step-by-step in Fig. S3. Without such a treatment, as it was the
case in ref. \cite{NarangJCIS2017}, the behavior of the
autocorrelation function at small times has no relation with the
diffusion properties of the suspended particles.

A cumulant analysis at each temperature, using the same criteria,
was made in both suspensions only for the contribution to the
autocorrelation function due to the Brownian motion of microgel
particles. More details about the procedure are reported in the SI.
The temperature dependence of $R_{h}$ is shown in Fig. S4. The
results for the suspension in pure water were validated by the
cumulant analysis performed by means of the Brookhaven commercial
software.

   \subsection{Raman spectroscopy experiments}

RS measurements were performed with a resolution of 5 cm$^{-1}$, by
using the 532 nm emission of a solid state laser (100 mW on the
sample). A backscattering geometry was realized using the 50x long
working distance objective of an Olympus BX40 microscope. The
scattered light was analyzed by an iHR320 imaging spectrometer
Horiba Jobin-Yvon, equipped with a Syncerity CCD camera. The signal
was dispersed by a 1800 grooves/mm grating that allowed acquisition
in the 750-1800 cm$^{-1}$ spectral range. Microgel samples in water
and in water-trehalose solution were measured at three temperatures
(283, 300 and 318 K) chosen in relation to the volume phase
transition temperature. At 283 K both samples are in swollen
condition; at 318 K both samples have collapsed; at 300 K microgels
in water and in water-trehalose are, respectively, below and above
the transition. The temperature was controlled by using a FTIR600
Stage by Linkam Scientific Instruments, equipped with a Linkam pump
system using liquid nitrogen as coolant. Spectra were collected by
cumulating several repetitions, corresponding to 2 hours of
acquisition at each temperature. Spectra of the two solvents were
also acquired. At each temperature, the solvent spectrum was
subtracted from that of the corresponding microgel sample, as much
as to avoid negative differences in the resulting `solvent-free'
spectrum. Notably, in the trehalose-containing system this operation
resulted in the complete removal of the peak at about 800 cm$^{-1}$,
the only trehalose signal with no superposition with those of
PNIPAM. More details about the solvent subtraction procedure are
provided in Figs. S5 and S6.

   \subsection{Differential scanning calorimetry measurements}

The volume phase transition of microgels in water and in
water-trehalose solution was analyzed by using a differential
scanning calorimeter SII DSC 7020 EXSTAR Seiko, equipped with liquid
nitrogen as cooling agent. The instrument was calibrated with
Indium, Zinc and heptane as standards. For each sample, 20--25 mg of
microgel suspension were hermetically sealed in a 60 $\mu$L
stainless steel pan equipped with rubber O-ring for operative
pressure up to 24 bar, taking care not to exceed the transition
temperature before measurement. The thermal analysis was carried out
on heating the sample from 0 to 60 $^{\circ}$C, using a scan rate of
5 $^{\circ}$C/min.

    \subsection{Molecular dynamics simulations}

        \subsubsection{Model and simulation procedure}

All-atom MD simulations were performed on a PNIPAM linear chain, in
diluted condition, both in water and in water-trehalose mixture at
the same trehalose mole fraction ($x_{tr}$=0.015) as used in the
experiments. The polymer chain was a 30-mer in atactic
configuration. PNIPAM was described with the OPLS-AA force field
\cite{JorgensenJACS1996} in the implementation of Siu et al.
\cite{SiuJCTC2012}, and trehalose with the OPLS-AA carbohydrates
force field \cite{DammJCC1997}. The model for water was TIP4P/ICE
\cite{AbascalJCP2005}, as it proved to properly reproduce the PNIPAM
behavior in water over a wide temperature range, including the
coil-globule transition \cite{TavagnaccoPCCP2018} and the dynamical
transition at low temperature \cite{ZanattaSCIADV2018,
TavagnaccoJPCL2019, TavagnaccoPRR2021}. The model of diluted
solution was built by adding randomly, around the polymer chain in
an extended conformation, 22849 water molecules for the PNIPAM-water
system, and 15598 water molecules plus 239 trehalose molecules for
PNIPAM in water-trehalose. Both systems were simulated at 283 and
318 K, the same temperatures as in the RS experiments, located below
and above the microgel volume phase transition in both solvents.
After equilibration, the mass density of the water-trehalose mixture
at 283 and 318 K was, respectively, $1092\pm0.02$ kg m$^{-3}$ and
$1086\pm0.01$ kg m$^{-3}$, consistent with the measured values of
1097 kg m$^{-3}$ and 1085 kg m$^{-3}$ at the same temperatures. MD
simulations were carried out in the NPT ensemble for 365 ns, using
the same protocol as in ref. \cite{TavagnaccoPCCP2018}. Trajectory
acquisition and analysis were carried out with the GROMACS software
package (version 2016.1) \cite{AbrahamSX2015}. The VMD software
\cite{HumphreyJMG1996} was used for graphical visualization. Further
details are reported in the SI.

        \subsubsection{Trajectory analysis}

The last 75 ns of each trajectory were considered for analysis, by
sampling 1 frame every 5 ps. The radius of gyration of the polymer
chain was calculated as $R_g = \sqrt{{\sum_i m_i s_i^2 / \sum_i
m_i}}$, where $m_i$ is the mass of the $i$-th atom and $s_i$ its
distance from the center of mass of the chain. The water accessible
surface area (WASA) of PNIPAM is defined as the surface of closest
approach of water molecules to the solute molecule, where both
solute and solvent are described as hard spheres. Numerically, this
quantity was calculated as the van der Waals envelope of the solute
molecule extended by the radius of the solvent sphere about each
solute atom center. A spherical probe of radius 0.14 nm was used and
the values of the van der Waals radii were taken from the literature
\cite{BondiJPC1964, EisenhaberCC1995}. The distributions of WASA
values were calculated with a bin of 0.1 nm$^2$. Water molecules in
the first hydration shell of PNIPAM were defined as having the
oxygen atom at a distance from oxygen, nitrogen, or methyl carbon
atoms of PNIPAM lower than the first minimum distance of the
corresponding radial distribution function. This cutoff value
corresponds to 0.35 nm for nitrogen and oxygen atoms and to 0.55 nm
for methyl carbon atoms. Water molecules in the first hydration
shell of trehalose were defined as having their oxygen within 0.35
nm from a trehalose oxygen atom. Finally, trehalose molecules in the
first solvation shell of PNIPAM were defined as having their
anomeric oxygen within 0.6 nm from oxygen, nitrogen or methyl carbon
atoms of PNIPAM. As the anomeric oxygen is located in the center of
the trehalose molecule, this criterion entails the proximity to
PNIPAM also of other sugar moieties, thereby selecting trehalose
molecules with a significant interaction with PNIPAM. The bulk
solvent properties were evaluated for molecules at a distance
greater than 3 nm from the polymer.

The trehalose concentration as a function of the distance from the
PNIPAM surface was evaluated by calculating the trehalose mole
fraction $x_{tr}$, given by the ratio between the anomeric oxygen
atoms of trehalose and the sum of the trehalose anomeric and water
oxygen atoms, in shells 0.5 nm thick. The number of PNIPAM-PNIPAM,
PNIPAM-water, PNIPAM-trehalose and water-trehalose hydrogen bonds
(HBs) was also calculated. The hydrogen bonding interaction was
identified by adopting the geometrical criteria of an acceptor-donor
distance (A$\cdots$D) lower than 0.35 nm and an angle $\Theta$
(A$\cdots$D--H) lower than 30$^{\circ}$, irrespective of the AD
pair. The lifetime of PNIPAM-water HBs, $\tau_{PW-HB}$, was
estimated by calculating the intermittent HB autocorrelation
function and by taking the time at which its amplitude decreased to
1/e. The exchange time of water (trehalose) from the PNIPAM first
hydration (solvation) shell to the bulk solution, $\tau_{FHS}$, was
estimated from the time evolution of the number fraction of water
(trehalose) molecules in the first shell, by taking the time at
which the corresponding autocorrelation function decayed to 1/e of
its amplitude.

The translational diffusion coefficient of PNIPAM and trehalose
hydrogen atoms and of bulk water oxygen atoms was calculated from
the long-time behavior of their mean square displacement (MSD), as
$D=\frac{1}{6}\lim_{t\to \infty} {d \over dt} \langle
|\mathbf{r}(t)-\mathbf{r}(0)|^2 \rangle$, where $\mathbf{r}(t)$ and
$\mathbf{r}(0)$ respectively represent the position vector of the
atom at time $t$ and $0$, and $\langle \cdots\rangle$ denotes an
ensemble and time origin average. In addition, we also calculated
the MSD of the oxygen atoms of PNIPAM and trehalose hydration water,
in a time window of 500 ps.

\section{\label{ResDisc}Results and Discussion}

     \subsection{\label{DLS}Dynamic light scattering}

\begin{figure}
\includegraphics[width=9.5 cm]{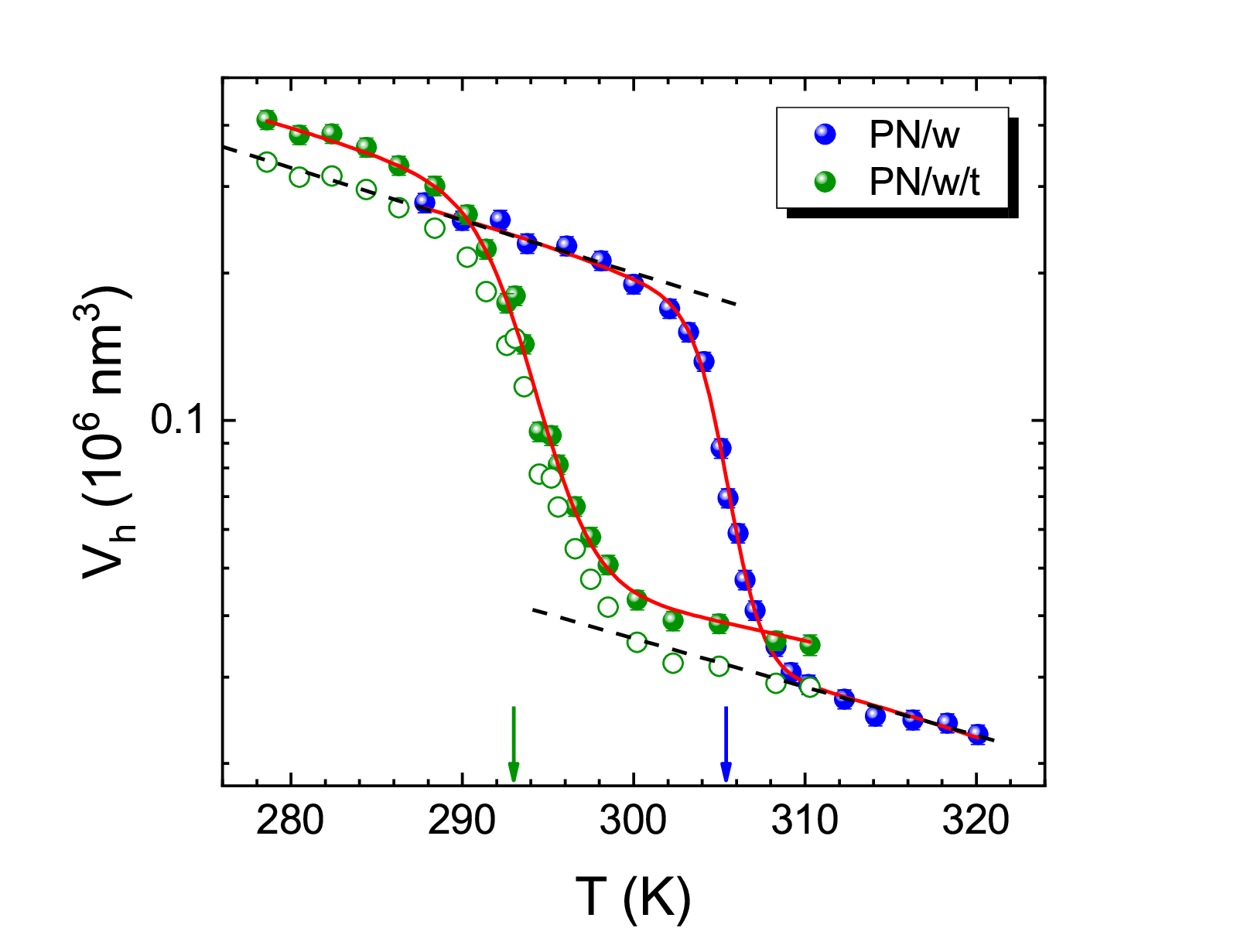}
\caption{Hydrodynamic volume, $V_{h}$, on a logarithmic scale,
plotted as a function of temperature, $T$, for PNIPAM microgels in
water (PN/w) and in water-trehalose solution (PN/w/t). Solid lines
are obtained by fitting the data with the two-state transition model
(equation (\ref{eq:vH})); arrows indicate the transition temperature
$T_{c}$. Open circles are obtained by dividing by a factor 1.2 the
data in water-trehalose. Dashed straight lines highlight that in
trehalose the particle volume is a 20$\%$ higher than in pure water
at the same temperature.} \label{fig:V(T)}
\end{figure}

DLS experiments revealed that the size of PNIPAM microgels, measured
by the hydrodynamic volume $V_{h}=\frac{4}{3}\pi R_{h}^{3}$, is
affected by the solvent medium in which the particles are suspended,
and allowed us to monitor the size change induced by temperature
variations in water and in water-trehalose solution (Fig.
\ref{fig:V(T)}). Upon heating, the value of $V_{h}$ has a sharp drop
within a narrow temperature range, revealing the occurrence of the
volume phase transition in both solvents. From a qualitative
comparison, two main differences with respect to the pure water
suspension emerge in the presence of trehalose at the investigated
concentration. Namely, the transition occurs at a temperature lower
more than 10 K, and the particles in their thermodynamically stable
state remain sensibly more expanded, both below and above the
transition.

Going into a deeper, yet model-independent evaluation, we find that
the increase of size in water-trehalose is proportional to the size
the microgel would have in water at the same temperature, the
difference being $\sim 20\%$ both in the swollen and the collapsed
state. In this respect, we note that the ratio between density of
water-trehalose and pure water (Fig. S1), which is constant over the
temperature range investigated, implies a constant ratio of about
1.18 between the volume of these solvents taken in quantities
containing the same amount of water. Moreover, since the molecular
volume of trehalose and water are about 350 {\AA}$^{3}$
\cite{GallinaJCP2010} and 30 {\AA}$^{3}$, respectively, one
trehalose molecule approximately occupies 18$\%$ the volume of 65.3
water molecules, 1:65.3 being the trehalose to water mole ratio in
the binary solvent. Thus, the $\sim 20\%$ greater hydrodynamic
volume detected when microgels are immersed in water-trehalose
solution could be well explained by assuming a water-cosolute
ability to enter/exit the microgel particles across the volume phase
transition maintaining the same molar composition as the external
medium. This rough estimate also suggests that the amount of water
permeating the polymer network, in the swollen or collapsed state,
is independent of the presence of trehalose. Our observation, made
on the microgel colloidal scale, is in agreement with the results of
isothermal titration microcalorimetry for PNIPAM chains
\cite{ShpigelmanJPSB2008}, suggesting that trehalose in the presence
of PNIPAM remains preferentially hydrated and mainly develops
water-mediated interactions with the polymer.

To obtain more quantitative information on the trehalose effect, we
derived the thermodynamic parameters associated to the microgel
transition, by applying a van't Hoff model to the temperature
dependence of $V_{h}$. In analogy to protein folding-unfolding, the
model considers a microgel as consisting of independent `cooperative
units', and the transition in each of these units as a
temperature-induced all-or-none transition from one state (swollen
state of the cooperative unit) to another state (collapsed state of
the cooperative unit) \cite{TiktopuloMACRO1994}. At any temperature
of the \textit{swollen}$\leftrightarrows$\textit{collapsed}
thermodynamic equilibrium, the measured value of $V_{h}$ is the
value averaged over the two cooperative unit populations, i.e.
$V_h(T)= V_{s} f_s(T)+ V_{c} f_c(T) $, with $V_{s}$ and $V_{c}$ the
volume of fully swollen and fully collapsed particles (i.e.,
composed of cooperative units all in the swollen or collapsed
state), while $f_{s}$ and $f_{c}$ denote the fraction at equilibrium
of swollen and collapsed units per particle, with $f_{s}+f_{c}=1$.
The equilibrium constant of the reaction, $K_{eq}=f_{c}/f_{s}$, is
given by $K_{eq}=\exp(-\Delta G^{0}_{T}/RT)$, with $R$ the gas
constant, $T$ the absolute temperature, and $\Delta G^{0}_{T}=\Delta
H^{0}_{T}-T \Delta S^{0}_{T}$ the standard free-energy change
associated to the transition at temperature $T$. It is reasonable
\cite{PrivalovJMB1999, PersikovPS2004, FucinosPLOS2014} that the
enthalpy and entropy changes at temperature $T$ negligibly differ
from their values $\Delta H^{0}$ and $\Delta S^{0}$ in the middle
point of the transition, i.e., at temperature $T_{c}$ defined by the
condition $K_{eq}(T_{c})=1$. This condition is equivalent to $\Delta
G^{0}(T_{c})=0$, implying $\Delta S^{0}=\Delta H^{0}/T_{c}$.
Combining these relations, gives
\begin{equation}
V_h(T)=\frac{V_{s} + V_{c} \exp\left[\frac{\Delta
H^{0}}{R}\left(\frac{1}{T_{c}}-\frac{1}{T}\right)\right]}{1 +
\exp\left[\frac{\Delta
H^{0}}{R}\left(\frac{1}{T_{c}}-\frac{1}{T}\right)\right]}
\label{eq:vH}
\end{equation}
Due to negative thermal expansion of molecular NIPAM
\cite{FutscherSCIREP2017, LiuCHEMCOMM2018}, the microgel network
contracts on heating. This temperature dependence in the volume of
fully swollen and fully collapsed particles is taken into account by
introducing a thermal contraction coefficient in each state,
$\alpha_{s}$ and $\alpha_{c}$. Therefore, in equation (\ref{eq:vH}),
one has $V_{s}=V_{s}^{c}[1-\alpha_{s}(T-T_{c})]$ and
$V_{c}=V_{c}^{c}[1-\alpha_{c}(T-T_{c})]$, with $V_{s}^{c}\equiv
V_{s}(T_{c})$ and $V_{c}^{c}\equiv V_{c}(T_{c})$ the hydrodynamic
volume extrapolated to $T_{c}$ from both sides of the transition.
The fit of the experimental data with equation (\ref{eq:vH}), shown
in Fig. \ref{fig:V(T)}, indicates that the two-state transition
model describes very well the volume phase transition of microgels
in both solvent media. An equally good description is not provided
by other thermodynamic models, e.g. the Flory-Rehner theory
\cite{LopezSM2017}.

The resulting best fit parameters, reported in Table \ref{tab:fit},
can be used to characterize the microgel ability to change volume in
response to temperature variations. To this end, we considered the
thermal effect that occurs without change of thermodynamic state
separately from the effect due to the conformational transition.
When particles are far from the transition region, either in the
swollen or in the collapsed state, their contraction coefficient in
the two solvents is the same within experimental error, suggesting
that the presence of trehalose does not appreciably change the
elasticity of the polymer network, a property reflected into its
ability to contract. Although the value of $\alpha_{c}$ is
significantly lower than (about one half) that of $\alpha_{s}$,
however it remains different from zero, confirming a residual
capacity of microgels to change volume in the collapsed state
\cite{BischofbergerSCIREP2015, NinarelloMACRO2019}. On the other
hand, a direct measure of the microgel ability to shrink across the
conformational transition is provided by the ratio
$V_{s}^{c}/V_{c}^{c}$. Notice that this quantification differs from
that used in literature, where the ratio is calculated between the
particle hydrodynamic radius well below and well above $T_{c}$,
without separating the effect of thermal contraction. The values of
$V_{s}^{c}/V_{c}^{c}$, equal within the error, are $5.2\pm 0.4$ and
$5.7\pm 1.1$ in water and in water-trehalose solution, respectively.
These numerical results are consistent with our previous
model-independent observation of a constant ratio, below and above
the transition, of $V_h$ values in the two solvents, further
demonstrating that equation (\ref{eq:vH}) is suitable to describe
the data behavior.

\begin{table*}
\caption{\label{tab:fit} Thermodynamic parameters obtained by
fitting with equation (\ref{eq:vH}) the temperature-dependent
hydrodynamic volume $V_{h}$ of PNIPAM microgels in water (PN/w) and
in water-trehalose solution (PN/w/t): volume phase transition
temperature ($T_{c}$); enthalpy of transition ($\Delta H^0$) per
mole of cooperative units; hydrodynamic volume extrapolated to
$T_{c}$ from low temperatures below the transition ($V_{s}^{c}$) and
from high temperatures above the transition ($V_{c}^{c}$); thermal
contraction coefficient in the swollen ($\alpha_s$) and collapsed
state ($\alpha_c$).} \vspace{0.2 cm}
\centering{\scalebox{0.84}{\begin{tabular}{l c c c c c c} \hline
Sample & $T_c$  (K) & $\Delta H^0$  (kJ mol$^{-1}$) & $V_{s}^{c}$ (nm$^{3}$) & $V_{c}^{c}$ (nm$^{3}$) & $\alpha_s$ (K$^{-1}$)& $\alpha_c$  (K$^{-1}$)\\
\hline
PN/w & $304.8\pm0.1$ & $800\pm60$ & $(1.65\pm0.07) 10^{5}$ & $(3.2\pm0.1) 10^{4}$ & $(3.8\pm0.4) 10^{-2}$ & $(1.9\pm0.2) 10^{-2}$ \\
PN/w/t & $293.0\pm0.3$ & $430\pm50$ & $(2.7\pm0.3) 10^{5}$ & $(4.7\pm0.4) 10^{4}$ & $(3.5\pm0.9) 10^{-2}$ & $(1.4\pm0.6) 10^{-2}$ \\
\hline
\end{tabular}
}}
\end{table*}

According to the parameters in Table \ref{tab:fit}, the
swelling-to-collapse transition is an endothermic process ($\Delta
H^0>0$) and also occurs with an increase of entropy ($\Delta
S^0=\Delta H^0/T_{c}>0$). Although a drastic decrease of both
$\Delta H^0$ and $\Delta S^0$ is observed in the presence of
trehalose, a transition temperature $T_c$ lower by almost 12 K stems
from a lower transition enthalpy which is not compensated by the
concomitant transition entropy reduction. This highlights a
remarkable effectiveness of trehalose in stabilizing the collapsed
state of PNIPAM microgels. Since, according to the model, the
enthalpy and entropy changes refer to one mole of cooperative units
undergoing the transition, to derive microscopic information from
these parameters it is necessary to calculate the number of monomers
per cooperative unit, in the two samples. This is done by dividing
$\Delta H^0$ by the transition enthalpy per mole of PNIPAM residues,
a quantity obtained from calorimetry. In this respect, we note that
several calorimetric studies have measured this quantity for PNIPAM
in water \cite{AnderssonJPSB2006, WoodwardEPJ2000,
BischofbergerSCIREP2014, SchildJPC1990, TiktopuloMACRO1994,
KujaraMACRO2001, FujishigeJPC1989, DingMACRO2005, SunSM2013}, but no
determination is available in water-trehalose solution at the
present concentration. Our thermal analysis provided $4.8\pm 0.5$ kJ
mol$^{-1}$ in water and $3.1\pm 0.3$ kJ mol$^{-1}$ in
water-trehalose. Notice that the first of these values is in
agreement with the numerous literature data, whose average is
$5.6\pm 0.8$ kJ mol$^{-1}$, and in particular is very close to the
values reported for microgels prepared with the same synthesis
protocol used in the present study \cite{AnderssonJPSB2006,
WoodwardEPJ2000}. With these values, the PNIPAM residues per
cooperative unit in water and in water-trehalose are $170\pm 30$ and
$140\pm 30$, respectively. This suggests a slight reduction of
cooperativity for the transition in the presence of trehalose,
despite the uncertainty of our estimate. These results for PNIPAM
microgels are reminiscent of results for proteins, where additives
can modulate the extent of folding cooperativity, and a less
cooperative process is found to be associated to an increased
protein stability \cite{JethvaJPCB2017, JethvaBIOCHEM2018}. Thus, a
less cooperative conformational transition in the presence of
trehalose further reflects the stabilizing action of this sugar.

Per mole of PNIPAM residues, the increase of entropy at the microgel
collapse in water and in water-trehalose is $15.7\pm 1.6$ and
$10.6\pm 1.0$ J K$^{-1}$ mol$^{-1}$, respectively, with a
trehalose-induced reduction of 4.9 J K$^{-1}$ mol$^{-1}$. The
reduction of heat absorbed, however, is 1.7 kJ mol$^{-1}$ only, a
small value compared to the typical energy of hydrogen bonding. This
is an indication that such a value reflects an energy balance
resulting from variations of several interactions. Indeed, $\Delta
H^0$ and $\Delta S^0$ provide global information about the
conformational transition, whose details at a microscopic level
remain hidden. As a demonstration, we may compare the cases of
aqueous PNIPAM solution with trehalose and ethanol, the latter being
an alcohol that directly exposes the hydrophobic moiety to the
surrounding. Despite the difference of the two additives, it is
experimentally observed \cite{BischofbergerSCIREP2014} that there
exists a water-ethanol composition ($x_{et}=0.063$) at which the
transition of PNIPAM exhibits similar thermodynamic parameters as
those found at the present trehalose concentration ($x_{tr}=0.015$),
namely, $T_c \sim 293$ K and an enthalpy change per mole of PNIPAM
residues of $\sim 3$ kJ mol$^{-1}$. Therefore, in order to derive a
clear microscopic picture of the mechanism of transition of PNIPAM
in the presence of trehalose, analyses at the molecular scale
constitute an essential further step of investigation, as reported
in the next sections.

    \subsection{\label{RS}Raman spectroscopy}

\begin{figure}
\includegraphics[width=8.5 cm]{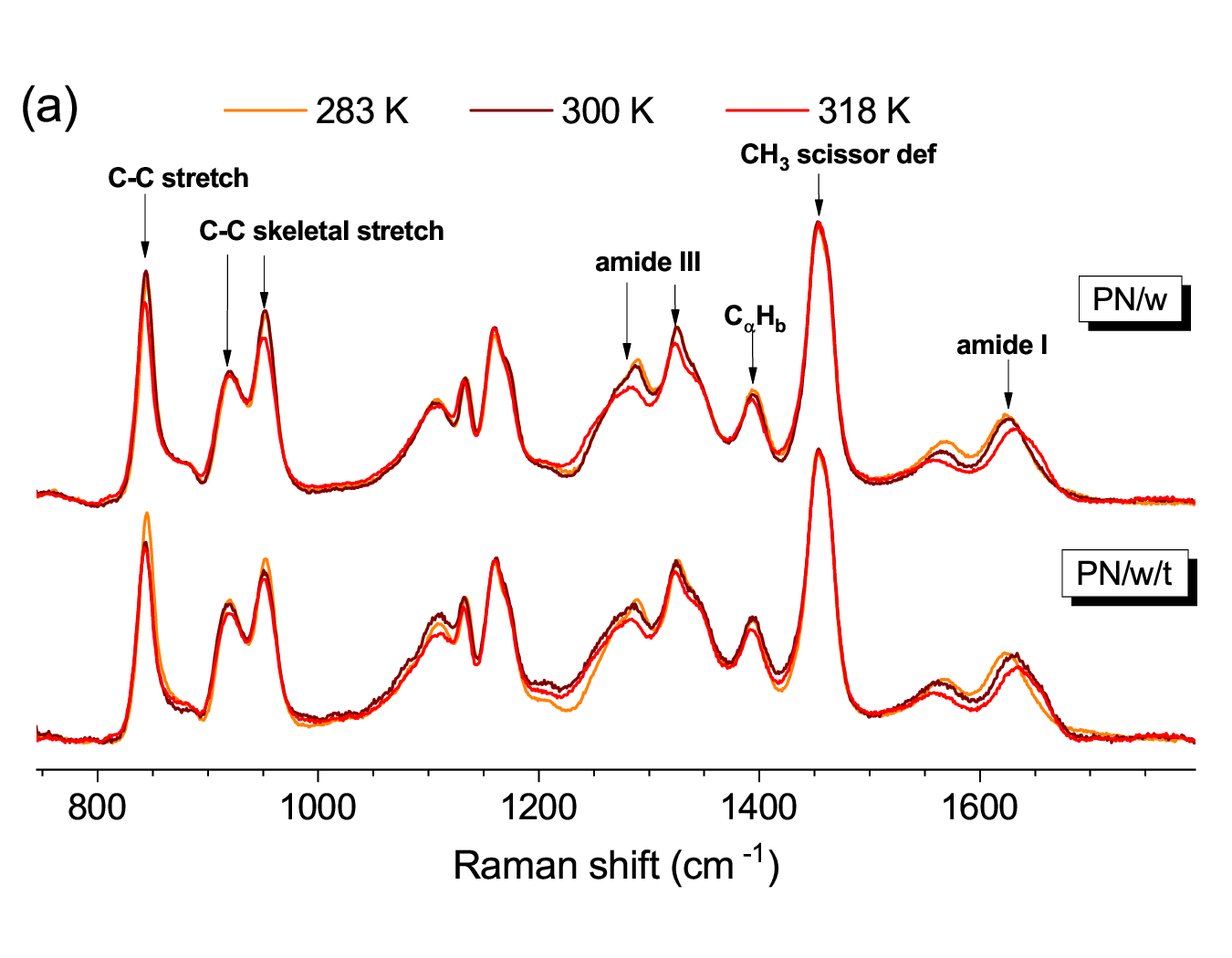}
\includegraphics[width=8.5 cm]{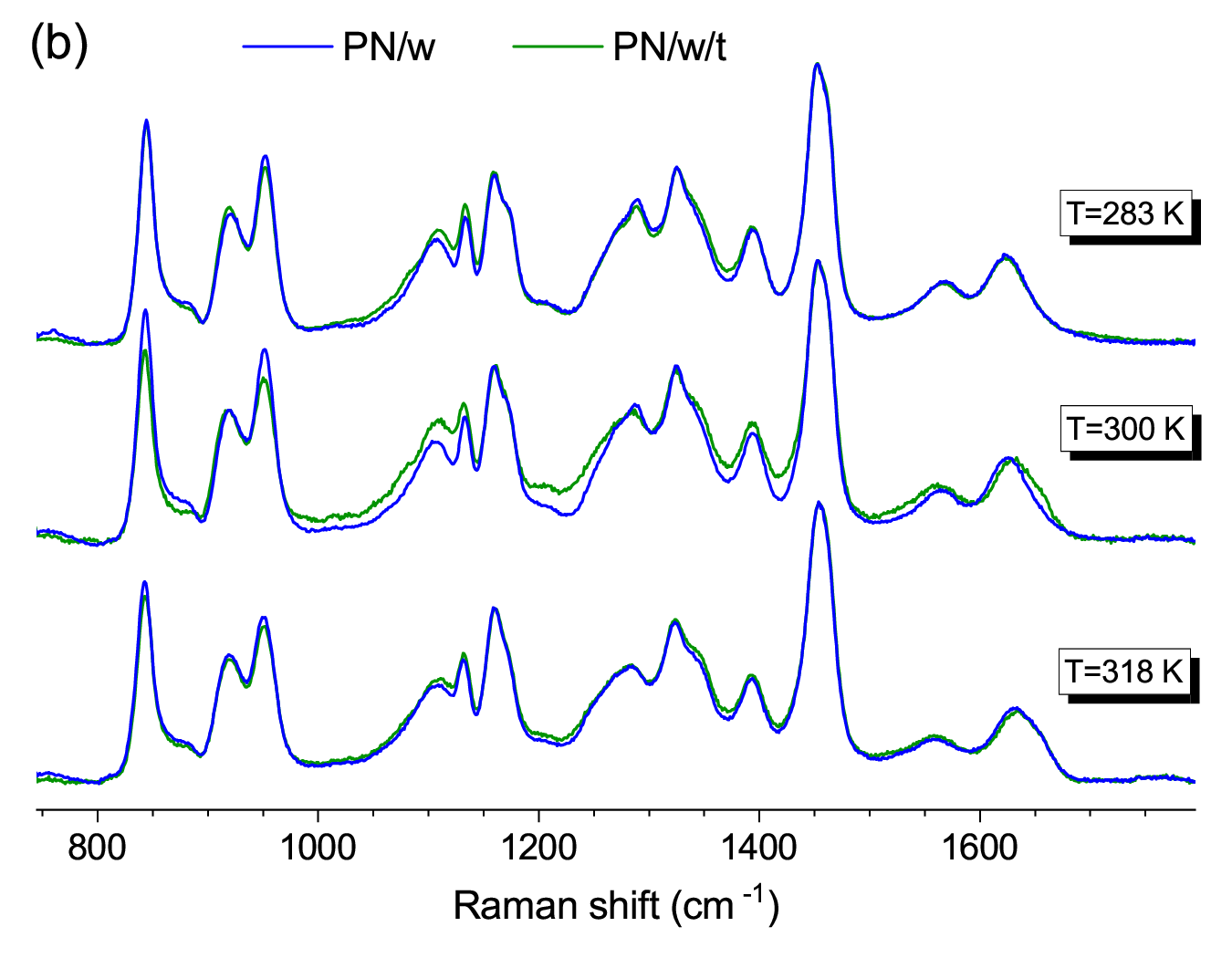}
\caption{(a) Raman solvent-free spectra of PNIPAM microgels in water
(PN/w) and in water-trehalose solution (PN/w/t) at different
temperatures, as indicated in the legend. The assignment of the
principal vibrational modes of PNIPAM are indicated by arrows. (b)
Comparison between the solvent-free spectra of PNIPAM microgels in
water and in water-trehalose solution at each temperature. In both
panels, the spectra are presented normalized to the intensity of the
isopropyl group's CH$_{3}$ deformation band at $\approx 1454$
cm$^{-1}$.} \label{fig:cfr_sub}
\end{figure}

\begin{figure}
\begin{center}
\includegraphics[width=8.5 cm]{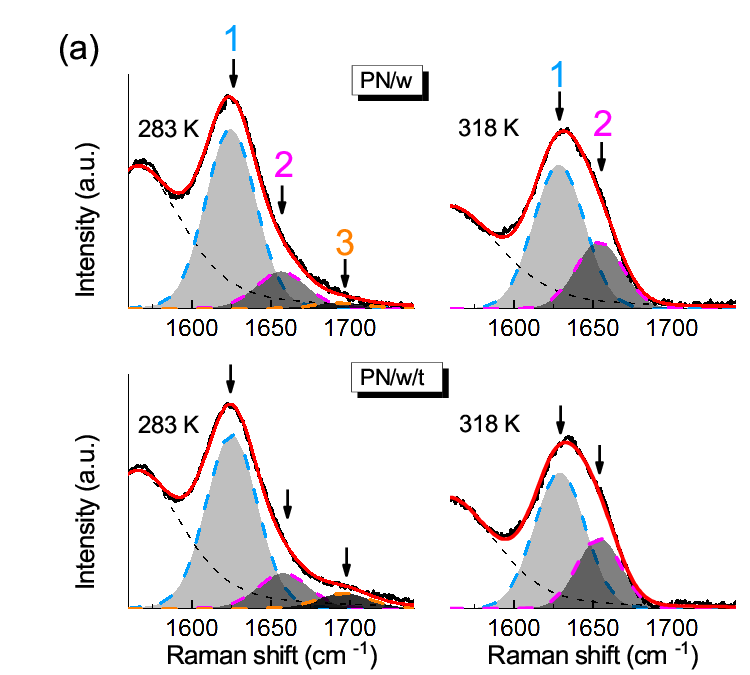}
\includegraphics[width=6.5 cm]{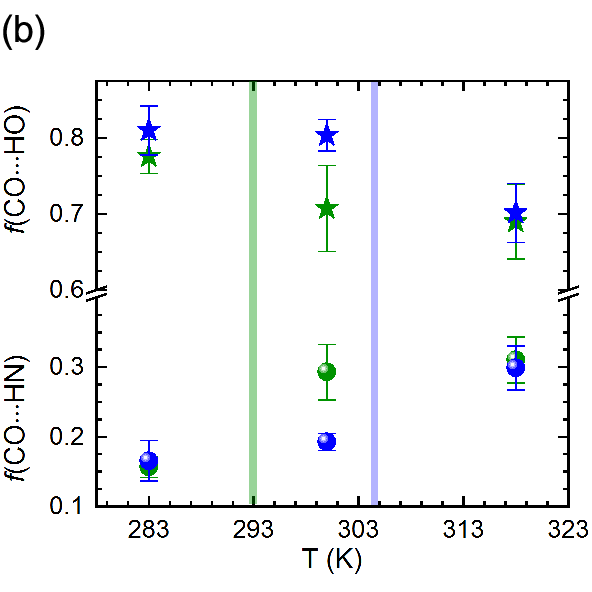}
\caption{(a) Fit of Raman solvent-free spectra in the amide I region
for PNIPAM microgels in water (PN/w) and in water-trehalose solution
(PN/w/t), below ($T=283$ K) and above ($T=318$ K) the volume phase
transition. The amide I band is decomposed into low (sub-band 1),
intermediate (sub-band 2) and high-frequency (sub-band 3)
pseudo-Voigt components. The three components have the following
parameters, the same for both samples and independent of
temperature: center $1627\pm 2$, $1656\pm 2$, $1695\pm 1$ cm$^{-1}$;
full-width-at-half-maximum $36\pm 3$ cm$^{-1}$. An additional
pseudo-Voigt function (black dashed line) is used to reproduce the
adjacent band around 1556 cm$^{-1}$. The red solid line represents
the total fitting curve. (b) Mole fraction of C=O groups involved in
intra-PNIPAM HBs, $f(C=O\cdots HN)$, and fraction of those involved
in HBs with hydroxyl groups, $f(C=O\cdots HO)$, for microgels in
pure water (blue) and in water-trehalose solution (green), plotted
against temperature. Colored bars indicate the corresponding values
of transition temperature $T_{c}$.} \label{fig:fit_Am}
\end{center}
\end{figure}

\begin{figure*}[h!]
\centering
\includegraphics[width=13 cm]{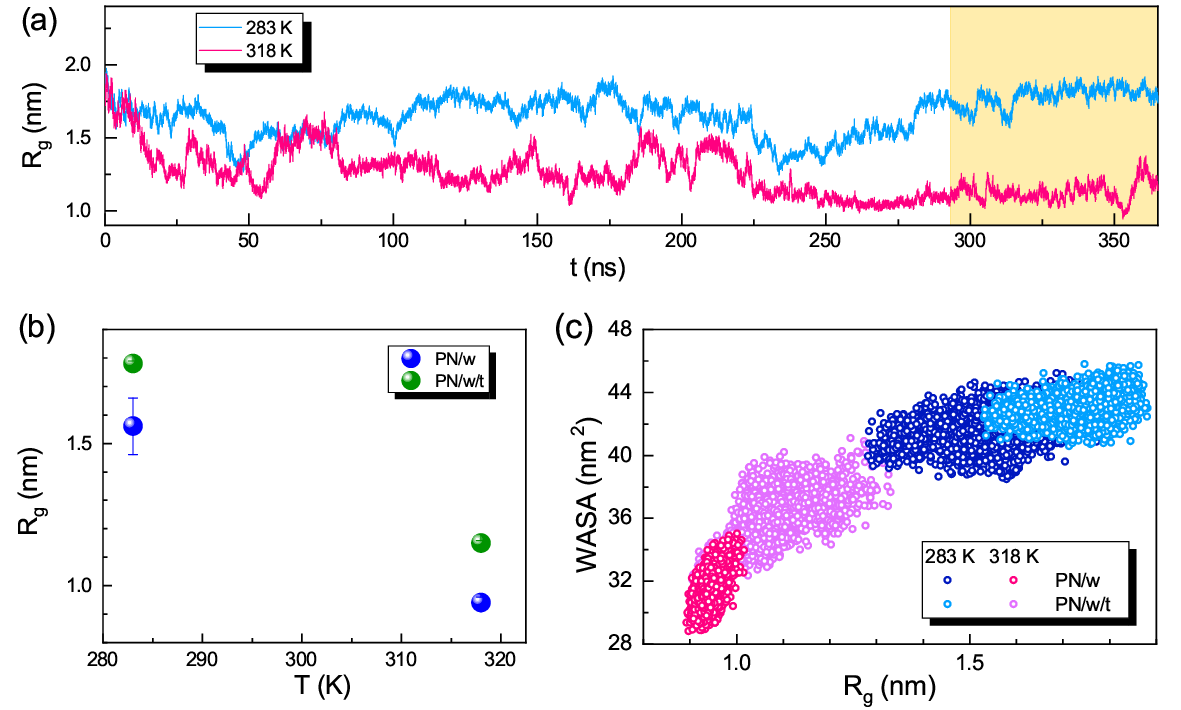}
\caption{(a) Time evolution of the radius of gyration, $R_g$, of the
studied PNIPAM chain in water-trehalose at 283 and 318 K. The time
interval used to calculate averaged properties (last 75 ns of each
trajectory) is highlighted in yellow. (b) Average radius of gyration
as a function of temperature $T$ for PNIPAM in water (PN/w) and in
water-trehalose (PN/w/t). (c) Correlation between WASA and radius of
gyration.} \label{fig:conf}
\end{figure*}

RS experiments were used to get insight into the solvation pattern
and intramolecular bonding of PNIPAM microgels in the two solvents,
at a molecular level. Fig. \ref{fig:cfr_sub}a shows the effect of
temperature, across the volume phase transition, on the solvent-free
spectra of each sample. Fig. \ref{fig:cfr_sub}b compares the
solvent-free spectra of the two samples at each studied temperature.
In the frequency range investigated, the amide I band conveys the
most relevant information, as it mainly arises ($\sim 80\%$) from
the C=O stretching vibration with only a minor contribution ($\sim
20\%$) from the C-N stretching mode. In each of the two samples,
this band shifts to higher frequency when microgels change from
swollen to collapsed (Fig. \ref{fig:cfr_sub}a), however this occurs
in a way that no difference in the band profile is observed when
microgels are both swollen (at 283 K) or when they are both
collapsed (at 318 K). This is shown in Fig. \ref{fig:cfr_sub}b.
Conversely, a band shift is observed at 300 K, when microgels are in
a different state. Notably, the solvent-free spectra of the two
samples, at 283 and 318 K, are well superimposed over the whole
frequency range, revealing that the signals of solvent and polymer
are not frequency shifted within the experimental sensitivity.
Therefore, direct binding of PNIPAM to trehalose is negligible, or
rather, is not distinguishable from that of PNIPAM to water.

Quantitative information was obtained from the curve fitting of the
spectra in the 1530-1740 cm$^{-1}$ frequency range, including the
amide I band. This latter is well reproduced by assuming the
presence of three sub-bands, each one described by a pseudo-Voigt
function, which are constant in position and shape and only change
in intensity upon temperature variation. The low-frequency component
($\approx 1627$ cm$^{-1}$) is attributed to C=O groups hydrogen
bonded to OH groups, the intermediate-frequency one ($\approx 1656$
cm$^{-1}$) to C=O groups intramolecularly hydrogen bonded to the NH
group \cite{MaedaLANG2000}, and the high-frequency component
($\approx 1695$ cm$^{-1}$) is ascribed to unbonded C=O groups
\cite{AhmedJPCB2009}. This testifies to the high sensitivity of the
amide I band to changes in both PNIPAM solvation and intra-PNIPAM
interactions. In particular, in the presence of trehalose, the
low-frequency component is assumed dominated by the interaction with
water, according to the experimental suggestions discussed in sec.
\ref{DLS} and the simulation results reported in the next sec.
\ref{MD}. The observed frequency upshift across the transition
reflects a change in the intensity balance among the three
sub-bands, with a higher relative increase of the
intermediate-frequency contribution. An additional pseudo-Voigt term
has been used to describe the band around 1565 cm$^{-1}$, that
partially overlaps the amide I band. The total fit curves together
with their individual contributions are displayed in Fig.
\ref{fig:fit_Am}a.

Assuming that the different C=O species have the same Raman cross
section, as found in poly-amide systems \cite{AhmedJPCB2009,
MaitiJACS2004}, we calculated the mole fraction of unbonded C=O as
the relative area of the high-frequency sub-band (number 3 in Fig.
\ref{fig:fit_Am}a), i.e. $f^{unbonded}=A_{1695}/A_{tot}$, with
$A_{tot}=A_{1695}+A_{1656}+ A_{1627}$ the sum of peak areas of the
three components. This small fraction is found to be temperature and
only slightly solvent dependent. Indeed, it vanishes in both samples
at 318 K, and at 283 K it is $\sim 4\%$ higher in the presence of
trehalose. However, this difference is not significant within the
experimental error, as demonstrated by the populations of the other
C=O species, hereafter calculated. The mole fraction of C=O involved
in intra-PNIPAM HBs is provided by the relative area $f(C=O\cdots
HN)=A_{1656}/A_{tot}$ of the intermediate-frequency sub-band (number
2 in Fig. \ref{fig:fit_Am}a), while the relative area $f(C=O\cdots
HO)=A_{1627}/A_{tot}$ of the low-frequency sub-band (number 1 in
Fig. \ref{fig:fit_Am}a) provides, within the assumption of
negligible bonding with trehalose, the fraction of hydrated C=O.
These quantities, plotted in Fig. \ref{fig:fit_Am}b as a function of
temperature, are coincident within the error for microgels in the
two solvents, below and above the transition, indicating that
neither the intramolecular bonding nor the hydrophilic hydration of
PNIPAM are affected appreciably by the presence of trehalose. These
findings support, with information at the molecular scale, the idea
suggested by the analysis of hydrodynamic volume data (see sec.
\ref{DLS}) that trehalose-induced effects are mostly water-mediated.

Fig. \ref{fig:fit_Am}b also reveals changes of interactions
associated to the microgel conformational transition. Even in the
swollen state a non-negligible amount of C=O groups, around 16$\%$,
is engaged in intramolecular bonding, and it increases in the
collapsed state, reaching about 30$\%$. Concerning hydration, most
(around 80$\%$) of the C=O groups are hydrogen bonded to water in
the swollen microgels, but only a small fraction, around 10$\%$, of
these groups dehydrate during the collapse transition. So, there is
still a lot of water retained in the collapsed particles. This
behavior favorably compares with other experimental and simulation
results for PNIPAM microgels \cite{AhmedJPCB2009} and linear chains
in water \cite{MaedaLANG2000, SunMACRO2008, PeltonJCIS2010,
TavagnaccoPCCP2018}. We note that a quantitative comparison with
literature data is prevented since several $f$ values are reported,
suggesting a possible effect due to the polymer topology and
concentration. The comparison in Fig. \ref{fig:fit_Am}b, however, is
not affected by this issue, as it refers to the same microgel
particles, at the same concentration.

    \subsection{\label{MD}Molecular dynamics simulations}

        \subsubsection{Structural properties}

To complement the microscopic information derived by RS experiments,
we rely on all-atom MD simulations of a PNIPAM 30-mer, mimicking a
linear segment of the microgel network. The time evolution of the
chain size (as quantified by the polymer radius of gyration $R_g$)
is reported in Fig. \ref{fig:conf}a for the trehalose-containing
system. The behavior shows that extended conformations are mainly
populated at 283 K, while the chain collapse has already occurred at
318 K. The average value of $R_g$ in the two solvents is compared in
Fig. \ref{fig:conf}b. While $R_g$ decreases from 283 to 318 K, the
chain size in the presence of trehalose is slightly bigger than in
pure water, both in the coil and the globule state. This difference
qualitatively matches the behavior of the hydrodynamic volume
observed by DLS in our PNIPAM microgels (see Fig. \ref{fig:V(T)}).
The water accessible surface area (WASA), a property strictly
correlated to the chain size in aqueous environment, confirms the
results for $R_g$, as shown in Fig. \ref{fig:conf}c. The associated
distributions of $R_g$ and WASA are reported in Figs. S7 and S8.

\begin{figure}
\includegraphics[width=7 cm]{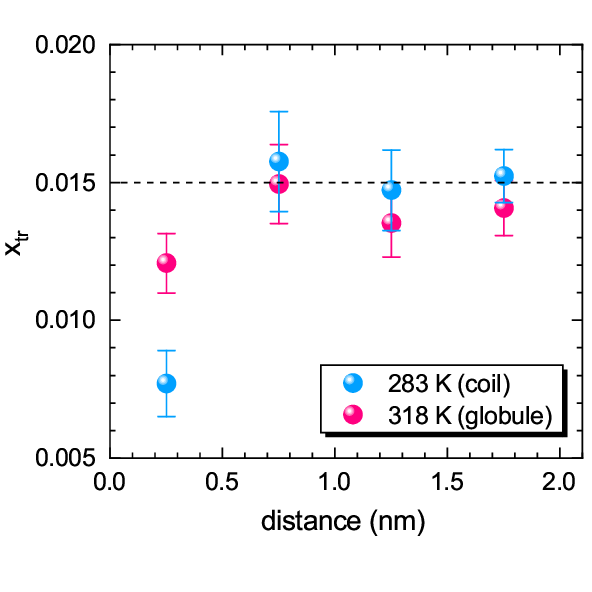}
\caption{Mole fraction of trehalose, $x_{tr}$, as a function of the
distance from the surface of the PNIPAM chain, in the coil (283 K)
and globule (318 K) conformation. A dashed horizontal line indicates
the value in the bulk solution.} \label{fig:Xt}
\end{figure}

The larger chain size in the presence of trehalose could derive from
a preferential adsorption of the sugar on the polymer, since
portions of the chain in a more extended conformation would favour
the interaction with trehalose through a greater accessible surface.
To verify whether this occurs, we calculated the local concentration
of trehalose as a function of the distance from the surface of the
PNIPAM chain (Fig. \ref{fig:Xt}). We find that the mole fraction of
trehalose in close proximity to the chain is lower than 0.015 found
in the bulk solution. This means that the polymer surface layer is
depleted of trehalose, and therefore preferential adsorption is
ruled out. Notably, this behavior is much more pronounced at 283 K,
in the coil conformation. These results for PNIPAM are quite
remarkable, since they are very similar to what observed for
globular proteins in solution, where trehalose is found to be
preferentially excluded from the protein surface
\cite{CourtenayBIOCHEM2000}. Indeed, an increased exclusion from the
vicinity of the denatured protein is proposed to be the physical
basis of the trehalose stabilizing action
\cite{TimasheffBIOCHEM1997, KaushikJBC2003}, a conjecture that seems
to be confirmed by the present findings.

\begin{figure*}
\begin{center}
\includegraphics[width=14 cm]{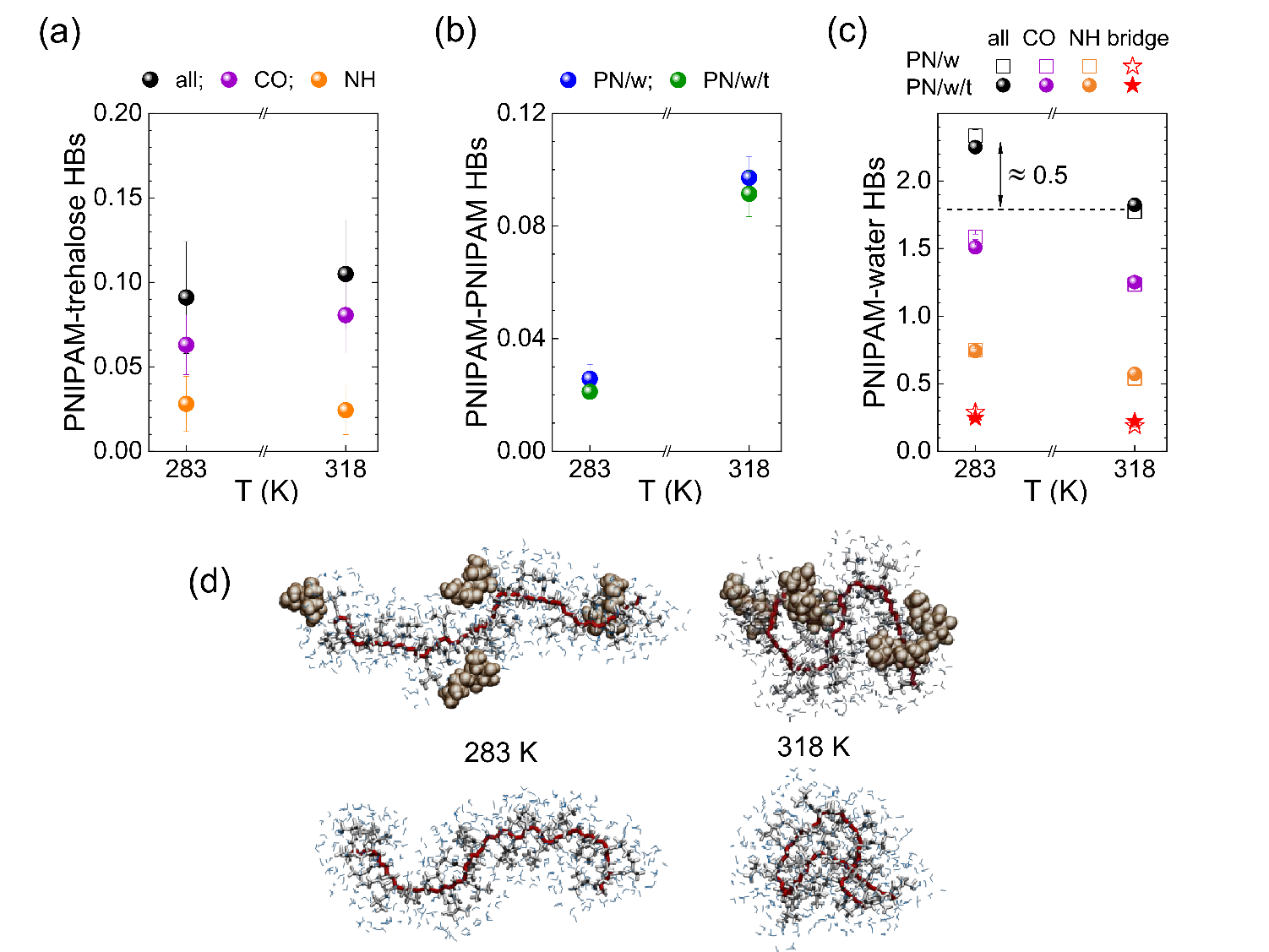}
\end{center}
\caption{Average number of (a) PNIPAM-trehalose, (b) PNIPAM-PNIPAM,
(c) PNIPAM-water HBs per PNIPAM repeating unit at 283 and 318 K. In
panels (a) and (c), the total number of HBs and its contributions
involving carbonyl (C=O) and amine (N-H) groups are shown, as
indicated in the legend. In panel (c), the number of water molecules
bridging two different amide groups of the polymer chain is also
reported. Error bars are within the symbol size. Panel (d) shows
snapshots of the simulated PNIPAM chain in water-trehalose (top) and
in water (bottom) at 283 and 318 K. PNIPAM backbone carbon atoms are
shown in red, hydrogen and side-chain atoms are represented in gray.
Water and trehalose molecules in the first solvation shell are
displayed in blue and yellow, respectively.} \label{fig:PN_wat}
\end{figure*}

To further investigate the trehalose arrangement, we looked for
trehalose molecules bound to hydrophilic groups of the polymer. On
average, only one out of ten PNIPAM residues forms HBs with
trehalose, irrespective of temperature (Fig. \ref{fig:PN_wat}a). The
extent of this interaction is negligible compared to that of the
polymer with water, providing support to the assumptions made in the
RS analysis of sec. \ref{RS}. However, since the size of a trehalose
molecule is about three or four times that of a PNIPAM residue, the
binding or the proximity of even very few molecules helps the
polymer chain to adopt more extended conformations than in pure
water, explaining a slightly larger size both in the coil and
globule state. This situation is visually displayed in Fig.
\ref{fig:PN_wat}d, showing representative configurations of the
simulated chain in the two solvents, along with the water and
trehalose molecules in the first solvation shell (see Methods and
Figs. S9, S10). Although the presence of trehalose in the PNIPAM
surrounding induces more extended conformations, Fig.
\ref{fig:PN_wat}b shows that this makes no difference for the number
of intra-PNIPAM HBs in the coil and collapsed chain. Moreover, the
increase of intramolecular bonds across the conformational
transition is in agreement with the behavior observed in the RS
experiments (see Fig. \ref{fig:fit_Am}b), though the fraction of
bonds obtained in the simulations is lower than that obtained
experimentally. As noted in sec. \ref{RS}, such discrepancy can be
ascribed to the much higher polymer concentration of the
experimental samples and to the network architecture of microgels,
both of which making contacts between amide groups more likely.

We then explored in detail the hydration pattern of PNIPAM in the
two solvents. The shape of the radial distribution functions between
water oxygens and PNIPAM atoms is very similar in water and in
water-trehalose, showing three distinct hydration shells (see Fig.
S10). For the polymer chain in the coil and globule state, we find
that the coordination number of PNIPAM atoms by water oxygens, which
is calculated by integration of the corresponding radial
distribution function, is the same in the two solvents up to
distances greater than the thickness of the first hydration shell
(see Fig. S11). Therefore, trehalose at the investigated
concentration does not inhibit PNIPAM hydration. This finding is
confirmed by the number of PNIPAM-water HBs, shown in Fig.
\ref{fig:PN_wat}c. We calculated this quantity separately for the
different hydrophilic (carbonyl and amine) groups of the polymer,
and also identified HBs where a water molecule bridges two different
amide groups. The results show that across the coil-globule
transition the chain maintains a high water affinity, with the loss
of only about 0.5 HBs per PNIPAM residue, as pointed out in previous
studies for linear chains \cite{MaedaLANG2000, SunMACRO2008,
PeltonJCIS2010, AhmedJPCB2009}. Most notably, the number of HBs for
each type of hydrophilic group is very similar in the two solvents,
indicating that trehalose not only preserves the overall polymer
hydration but also keeps unchanged the pattern of hydrogen bonding
with water. These findings are in agreement with the results of the
RS experiments (see Fig. \ref{fig:fit_Am}b).

The mentioned results also suggest that maintaining the same
hydration level as in pure water can be an additional driving force
which induces more extended conformations of the PNIPAM chain, and
allows for higher WASA values compensating for the presence of
trehalose in the polymer surrounding. Another consequence of these
results is that the decrease of transition enthalpy per mole of
polymer residues, measured by calorimetry in the presence of
trehalose, cannot be ascribed to a lower dehydration of amide
groups, but is likely due to a smaller molar partial enthalpy of
water in the bulk of the water-trehalose solution
\cite{MillerJPCB2000}, a factor which thermodynamically favors the
transition to the collapsed state \cite{BischofbergerSCIREP2014}.

\begin{table*}
\caption{\label{tab:MD} Solvation properties from MD simulations of
PNIPAM in water (w) and in water-trehalose mixture (w/t). $HB_{TW}$
is the average number of trehalose-water HBs and $\tau_{TW-HB}$
their characteristic lifetime; $D_P$, $D_T$, and $D_W$ are the
diffusion coefficient of PNIPAM, trehalose, and bulk water,
respectively; $\tau_{PW-HB}$ is the lifetime of PNIPAM-water HBs,
and $\tau_{FHS}$ is the exchange time of water in the first
hydration shell. $D$ values are within 5\%.} \vspace{0.2 cm}
\centering{\scalebox{0.82}{\begin{tabular}{c c c c c c c c c c c c}
\hline
T   & $HB_{TW}$ & $\tau_{TW-HB}$ & $D_P$ (w) & $D_P$ (w/t) & $D_T$ & $D_W$ (w) & $D_W$ (w/t) & $\tau_{PW-HB}$ (w) & $\tau_{PW-HB}$ (w/t) & $\tau_{FHS}$ (w) & $\tau_{FHS}$ (w/t) \\
(K) & & (ps) & $(cm^2 s^{-1})$ & $(cm^2 s^{-1})$ & $(cm^2 s^{-1})$ & $(cm^2 s^{-1})$ & $(cm^2 s^{-1})$ & (ps) & (ps) & (ps) & (ps) \\
\hline
283 & $15.5(\pm0.1)$ & $93(\pm3)$ & $4.5\cdot10^{-7}$  & $1.4\cdot 10^{-7}$ & $4.2\cdot10^{-7}$ & $0.62\cdot10^{-5}$ & $0.35\cdot10^{-5}$ & $260(\pm30)$ & $360(\pm10)$ & $490(\pm10)$ & $760(\pm10)$ \\
318 & $15.0(\pm0.1)$ & $23(\pm3)$ & $12\cdot 10^{-7}$ & $7.5\cdot 10^{-7}$ & $17\cdot10^{-7}$ & $1.8\cdot10^{-5}$ & $1.07\cdot10^{-5}$ & $67(\pm3)$ & $88(\pm3)$ & $135(\pm10)$ & $230(\pm10)$ \\
\hline
\end{tabular}
}}
\end{table*}

We then analyzed the hydration and aggregation properties of
trehalose in our model. Indeed, the presence of hydrated-sugar
complexes acting as an effective cosolute has been proposed to
control the effect of carbohydrates on the PNIPAM LCST, in such a
way that the higher the sugar size and hydration, the larger the
decrease of the LCST, mainly for entropic effect
\cite{ShpigelmanJPSB2008}. We find a hydration number of
$18.0\pm0.1$, irrespective of temperature, in very good agreement
with the value of $\sim 20$ determined experimentally at the same
trehalose concentration \cite{LupiJPCB2012, FiorettoFOOD2013,
ComezJPCL2013}. On average, one trehalose molecule hydrogen bonds
$15.5\pm0.1$ and $15.0\pm0.1$ water molecules, respectively at 283
and 318 K (see Table \ref{tab:MD}), while it only forms 0.5
inter-trehalose HBs, which demonstrates a high water affinity and a
weak tendency to aggregate, as experimentally reported
\cite{LupiJPCB2012, FiorettoFOOD2013}. Particularly, we analyzed the
distribution of trehalose in the PNIPAM surrounding, looking for
some connectivity between sugar molecules residing in the first
solvation shell. We observe that these molecules never hydrogen bond
to each other, so that no clustering occurs at the polymer
interface. These findings, combined with the relatively long
trehalose-water HB lifetime, $\tau_{TW-HB}$ (see Table
\ref{tab:MD}), confirm the strong and preferential interaction of
trehalose with water and exclude the presence of clusters in the
polymer surrounding for the system here investigated.

        \subsubsection{Dynamical properties}

To complement these results, we studied the trehalose-induced effect
on the dynamics of the system constituents. To this end, we
monitored the MSD of PNIPAM and trehalose hydrogen atoms, and of
bulk water oxygen atoms (Fig. S12). All these atoms show a long-time
diffusive behavior so that it is possible to estimate their
diffusion coefficients, respectively $D_P$, $D_T$, and $D_w$. The
values are reported in Table \ref{tab:MD}. We observe that the
addition of trehalose induces a drastic slowdown of both PNIPAM and
water. In particular, the bulk water diffusivity decreases by almost
a factor of 2, at both temperatures. A similar reduction is observed
for PNIPAM at 318 K, but the effect is almost double at 283 K, where
$D_P$ is only $\sim 30\%$ of the one in pure water. We also note
that the temperature-induced change of PNIPAM diffusivity is in a
different relationship with the corresponding change of the bulk
solvent mobility. In fact, the value of $D_P$ in water roughly
scales with $D_w$, while in the trehalose-containing system the
reduction of $D_P$ from 318 to 283 K is greater than that of $D_T$
and $D_w$, suggesting that the presence of trehalose enhances the
slowing down of polymer local motions upon cooling. Concerning
hydration water, of both trehalose and PNIPAM, we observe that the
oxygen atoms never reach a diffusive regime within the studied time
interval, thus preventing an estimate of their diffusion
coefficient. However, the time behavior of the MSDs (Fig. S12) shows
that the hydration water dynamics is always slower than that of bulk
water. The molecules in the hydration shell of PNIPAM are slower
than those in the hydration shell of trehalose and, most
importantly, the mobility of PNIPAM hydration water is much lower in
water-trehalose solution. This is observed at 283 and 318 K, and
highlights the strong slowdown induced by trehalose also in the
dynamics of the polymer hydration water.

To better characterize the hydration shell dynamics, we calculated
three characteristic times, namely, the PNIPAM-water HB lifetime
($\tau_{PW-HB}$) and the exchange time of water and of trehalose in
the first hydration shell ($\tau_{FHS}$), as described in Materials
and Methods. At each temperature, $\tau_{FHS}$ of water molecules is
higher than $\tau_{PW-HB}$. This reflects the major constraints
imposed on these molecules by being hydrogen bonded rather than
residing in the polymer first hydration shell, and also depends on
their possibility to break HBs with amide groups without escaping
the surrounding of hydrophilic or hydrophobic PNIPAM domains. In the
presence of trehalose the value of these characteristic times
significantly increases. As shown in Table \ref{tab:MD}, the
PNIPAM-water HBs last $\sim 40\%$ longer, and the exchange of water
molecules from the hydration shell to the bulk needs $\sim 60\%$
more time. Moreover, the exchange of trehalose takes a time over one
order of magnitude longer than the exchange of water (see Fig. S13).
All these findings demonstrate that, independently of the polymer
conformation, trehalose provides an increased dynamical stability to
the polymer hydration shell. Again, these results for PNIPAM are
noteworthy, since they are closely reminiscent of the slowdown
caused by trehalose to protein local motions and hydration water,
helpful to obviate high-temperature damages in solution
\cite{CiceroneBIOPHYSJ2004, SubatraJPCB2015, MalferrariJPCL2016}.
Moreover, it should be pointed out the similarity with the mechanism
of protein cryoprotection described in a recent simulation study of
lysozyme in aqueous solution \cite{CamisascaJCP2020}. The study
revealed two structural relaxations of hydration water, with the
long-time process being sensitive to the protein structural
fluctuations and changing its temperature behavior in correspondence
to the protein dynamical transition. The effect of trehalose was to
strongly damp the fluctuations of the protein structure, and to
slowdown the long-time water relaxation to a much greater extent
than the other process. This effect was claimed to inhibit ice
nucleation and favor vitrification upon cooling, providing a
rationale to the cryoprotectant action of trehalose for proteins.
Our study reveals a very similar mechanism of slowdown induced by
trehalose, reflected in the long-lasting PNIPAM-water HBs combined
with a slow exchange of hydration-to-bulk water and an even slower
exchange of trehalose, all contributing to confine a small amount of
water molecules close to the polymer surface. Such molecules,
located within the polymer hydration shell, are necessarily most
prone to share hydrogen bonding interactions with polymer and
trehalose and are thus expected to be more slowed down and more
coupled to the structural fluctuations of the macromolecular system.
The analogy is reinforced by the similar time behavior of the MSD
observed for the hydration water that, in lysozyme and PNIPAM, never
reaches the diffusive regime at any temperature, in contrast to bulk
water.

In the simulation of lysozyme, the sugar-induced slowing down was
found to be associated with the presence, over the protein surface,
of transient trehalose patches formed by clusters of several
molecules \cite{CamisascaJCP2020}. In the PNIPAM system, however, no
clustering of trehalose is detected, whereas a clear slowing down of
both polymer and its hydration water is found, suggesting that a
massive binding of the sugar to the macromolecule is not an
essential ingredient. Since the absence of clustering in the system
here investigated might well be explained by the lower concentration
of trehalose (0.72 instead of 1.33 M), further investigations at
higher concentration would be useful to clarify the effect of
trehalose clusters at the polymer interface.

It is now important to stress that the hydration pattern of PNIPAM
and the localization of additive in its surrounding, are very
different in water-trehalose from those identified in water-ethanol
by another recent simulation study \cite{TavagnaccoJML2020}. Unlike
trehalose, ethanol preferentially adsorbs on the polymer surface,
mainly due to interactions between ethyl and isopropyl groups, but
also competes with water for hydrogen bonding to hydrophilic groups
and develops aggregates in the first solvation shell of the polymer.
Thus, important differences at the microscopic level characterize
the mechanisms governing the effect of these two additives on the
polymer transition, even in case of similar thermodynamic parameters
as discussed in sec. \ref{DLS}.

Overall, the simulations give us a microscopic picture of the effect
of trehalose on PNIPAM which indicates that, consistently with the
experimental results, the bioprotectant sugar preserves PNIPAM
hydration both below and above the transition, and stabilizes the
hydration shell by preferential exclusion and drastic slowdown of
local motions. These mechanisms are at odds with those put in place
by other additives, for instance ethanol, and are strikingly similar
to those observed in trehalose-protein systems.

\section{\label{Concl}Conclusions}

In this manuscript, we reported evidence that trehalose acts on
PNIPAM microgels through molecular mechanisms that are similar to
those governing its protective action against thermal denaturation
and cold damage of aqueous proteins. To this aim, we combined DLS
and RS measurements with MD simulations, to complement
single-particle information at the colloidal scale with atomic-scale
structural and dynamical information. From the temperature
dependence of the microgel hydrodynamic volume, which is well
described by the van't Hoff equation for the thermodynamic
equilibrium between the swollen state of the particles for $T<T_c$
and their collapsed state above it, the $T_c$ at the investigated
0.72 M trehalose concentration was found to be lowered by as much as
12 K, showing that trehalose promotes the collapsed conformation of
the microgel particles similarly to the way it stabilizes aqueous
proteins in their native state. Trehalose has proved to be even more
effective as a stabilizer of microgels than of proteins, as shown by
the exceptional reduction of transition temperature ($\Delta T_c$),
which is almost twice the increase of transition temperature in
proteins for the same sugar concentration
\cite{TimasheffBIOCHEM1997, KaushikJBC2003}.

In the presence of trehalose, microgels are found $\sim 20\%$ more
expanded than in pure water keeping unchanged their thermal
contraction, swelling capacity, and the amount of absorbed water,
with no change in composition of the solvent absorbed below and
above the transition. This suggests that sugar molecules remain
preferentially hydrated without directly interfering with the HBs
formed by the microgel network, thereby preserving polymer
hydration. The numerical simulations provided evidence, indeed, that
binding of some trehalose to the polymer occurs, but without
changing the water affinity for PNIPAM which, below and above the
transition, is expected to be preferentially hydrated.

Most crucially, we found an increased preferential exclusion of
trehalose in the swollen microgels, in close analogy to the
increased exclusion from the hydration shell of denatured proteins,
a mechanism thermodynamically explaining the sugar stabilizing
action \cite{CourtenayBIOCHEM2000, TimasheffBIOCHEM1997,
KaushikJBC2003}. The collapsed conformation of PNIPAM microgels is
not only thermodynamically but also kinetically stabilized, in
further analogy to protein behavior. Trehalose, in fact, was found
to deeply impact on the dynamics of the system by inducing a drastic
slowdown of the polymer local motions and of its hydration shell. A
reduced mobility, together with an increased residence time of water
in the polymer surrounding, is also one major mechanism explaining
the trehalose ability to counteract deleterious intramolecular
motions in delicate proteins and enzymes in solution, during
exposure to high temperatures \cite{CiceroneBIOPHYSJ2004,
SubatraJPCB2015, MalferrariJPCL2016}. Moreover, the long-lasting
PNIPAM-water interactions, and the reduced hydration-to-bulk
exchange of solvent molecules, contribute to confine a small amount
of slow water close to the polymer surface, suggesting similarities
with the protective mechanism of trehalose against cold damage in
proteins. In this respect, a natural and interesting extension of
this work will be to simulate a microgel network upon cooling to
ascertain the occurrence of a long-time structural relaxation in the
polymer hydration water, similar to the one detected in the recent
simulations of lysozyme in solution \cite{CorradiniSCIREP2013,
CamisascaJCP2020}, and to investigate how trehalose affects the
polymer dynamical transition, a phenomenon which is well established
for PNIPAM microgels only in water \cite{ZanattaSCIADV2018,
TavagnaccoJPCL2019}.

At this point, it is important to see the present study in the
context of the existing literature about the trehalose effect on the
PNIPAM stability in aqueous solution. In particular, the MD
simulation study by Narang and coworkers \cite{NarangJCIS2017}
claimed to explore the changes of PNIPAM intramolecular bonding and
solvent interactions across the coil-globule transition. Such
investigation was, however, limited to the effect produced by a
single trehalose molecule on one PNIPAM chain (35-mer) in water, a
system in which the trehalose concentration ($\sim$0.012 M) was
clearly unable to affect the polymer LCST with respect to that in
pure water. Moreover, the simulation was carried out at a single
temperature (300 K), which is still below the LCST in such a highly
diluted trehalose solution. Therefore, the results of ref.
\cite{NarangJCIS2017} only refer to trehalose-induced changes of
PNIPAM in the coil conformation, leaving unexplored those associated
to the coil-globule transition.

Another study that deserves a comment is the experimental work of
Shpigelman and coworkers \cite{ShpigelmanJPSB2008} focused on the
behavior of PNIPAM LCST at increasing concentration of different
sugars, including trehalose, as a way to indirectly probe the
stabilization effect of sugars on proteins. Results from isothermal
titration microcalorimetry, showing no preferential adsorption of
trehalose on the polymer, are found to be in agreement with our
findings. However, such study was based on the assumption that the
interactions between the two solvent components and the responsive
polymer reproduce those of a protein in the same aqueous medium.
This is a different approach to that adopted in the present work,
where the PNIPAM-trehalose interaction mechanism has been elucidated
at a molecular level and then compared to that observed in proteins.
In this respect, it should be stressed that the protein-analogue
behavior of PNIPAM is not to be taken for granted, as important
differences between the two macromolecular systems exist -- from the
complexity level of their chemical structure to the aggregation and
phase separation behavior, to the residual hydration degree of their
collapsed state -- which could be important in relation to specific
phenomena.

As a whole, our findings prove that PNIPAM microgels are capable of
efficiently mimicking the sensitivity of proteins to the
water-trehalose solvent, giving rise to thermodynamic and dynamic
mechanisms fully similar to those responsible for the sugar
bioprotective action. The results also highlight that these
mechanisms, for instance preferential exclusion, may strongly differ
from those, for example preferential adsorption, arising from the
polymer interaction with additives of opposite effect on proteins,
such as alcohols. Altogether, this work puts forward PNIPAM
microgels as a genuinely synthetic protein-like template, that will
be useful to investigate, under controlled conditions, the source of
different additive actions, a biological issue subject of intensive
research.

Future work will aim to provide generality to the present findings
for trehalose, by extending the investigation to other additives,
with different hydrophilicity and size, yet endowed with biological
function. Moreover, as trehalose is also the most effective sugar to
stabilize proteins during dehydration, PNIPAM microgels could be
used as models for a better understanding of the mechanisms required
to proteins to obviate this fundamentally different stress vector,
compared to freezing and heat denaturation \cite{CroweCRYO1990}.

\section*{Acknowledgments}
B.R., L.C, S.C. and C.P. acknowledge support from Universit\`{a} di
Perugia (``CarESS" project, D.R. n. 597); L.T., M.B., E.C. and E.Z.
from European Research Council - ERC (ERC-CoG-2015, Grant No. 681597
MIMIC); L.T., E.B., E.C. and E.Z. from Ministero dell'Istruzione,
dell'Universit\`{a} e della Ricerca - MIUR (FARE project R16XLE2X3L,
SOFTART). Support for computational time by CINECA-ISCRA grants
(Nos. HP10C1IX5O and HP10C9V0IP) is also acknowledged.

\clearpage
\newpage
\onecolumn
\setcounter{figure}{0}
\setcounter{section}{0}

\textbf{Supporting Information}\\
 \section{Experimental}

    \subsection{Analysis of DLS spectra with the cumulant method}

The quantity measured in a dynamic light scattering (DLS) experiment
is $G^{(2)}(q,t)\equiv \langle I(q,0)I(q,t) \rangle$, the
autocorrelation function of the scattered intensity. This is linked
to the autocorrelation function of the scattered field,
$G^{(1)}(q,t)\equiv \langle E^{*}(q,0)E(q,t)\rangle$, by the Siegert
relation $G^{(2)}(t)=A_{0}\left[1 + \beta |G^{(1)}(t)|^{2} \right]$,
where $A_{0}$ is a measured baseline and $\beta$ is the coherence
factor, an instrumental parameter of the order of unit. Therefore,
$[G^{(2)}(t)-A_{0}]^{1/2}$ is proportional to $G^{(1)}(t)$. For a
monodisperse diluted suspension of macromolecular objects,
$G^{(1)}(t)$ decays as a single exponential, so that
$\left[G^{(2)}(t)-A_{0}\right]^{1/2} \propto \exp(-\Gamma t)$, where
the decay time $\tau =\Gamma^{-1}$ is related to the translational
diffusion coefficient $D$ of the particles as $\tau=1/(D q^{2})$.
For a polydisperse suspension, $G^{(1)}(t)$ does not decay
exponentially, and its deviation from a single exponential behavior
contains information about the distribution of diffusion
coefficients. In this case $G^{(1)}(t)$, and hence
$\left[G^{(2)}(t)-A_{0}\right]^{1/2}$, can be represented as a
superposition of exponential contributions, each one arising from a
population of particles with a certain diffusion coefficient and
weighted by the population scattering intensity, i.e.,
\begin{equation}
G^{(1)}(t) =A \int_{0}^{\infty} I(\Gamma) \mathrm{e}^{-\Gamma t}
d\Gamma, \label{eq:cum}
\end{equation}
with $\Gamma=D q^{2}$, $I(\Gamma)$ the normalized distribution
function of the scattered intensity from particles with diffusion
coefficient $D$, and $A$ a proportionality constant. The quantity in
equation (\ref{eq:cum}) can be analyzed by the method of cumulants
\cite{BernePecora}, by expanding its logarithm in a power series in
$t$,
\begin{equation}
\ln G^{(1)}(t) = \ln A - K_1t+\frac{1}{2!}K_2t^2-\frac{1}{3!}K_3t^3
+ ... \label{eq:cum2}
\end{equation}
where $K_{n}$ is called the n$^{th}$ cumulant. The cumulants are
related to the moments of the diffusion coefficient distribution.
The explicit form of the first two of them is
\begin{eqnarray}
  K_{1} &=& \int_{0}^{\infty} I(\Gamma) \Gamma d\Gamma =<\Gamma>_{z}=q^{2}<D>_{z} \\
  K_{2} &=& \int_{0}^{\infty} I(\Gamma)
\left(\Gamma-<\Gamma>_{z}\right)^{2} d\Gamma =<\left(\delta
\Gamma\right)^{2}>_{z} =q^{4}<(\delta D)^{2}>_{z}
\end{eqnarray}
i.e., they respectively provide the average value and variance of
the diffusion coefficient distribution. The average values extracted
by this method, which are weighted by the particle scattering
intensity, are defined as z-average values, and denoted by
$<\cdots>_{z}$.\\
Therefore, by analyzing with a polynomial function the quantity
$\left[G^{(2)}(t)-A_{0}\right]^{1/2}\propto G^{(1)}(t)$ obtained
from the experiment, if the data are good at small times, one
obtains $<D>_{z}$ and its variance $<(\delta D)^{2}>_{z}$ from the
first and second-order polynomial terms. Then, the hydrodynamic
radius $R_{h}$ of the particles can be determined as the z-average
particle size by using the Stokes-Einstein relation
$R_{h}=k_{B}T/6\pi\eta <D>_{z}$. Here, $k_{B}$ is the Boltzmann's
constant, $T$ the absolute temperature, and $\eta$ the viscosity of
the dispersing medium.\\
In the presence of a scattering contribution due to the solvent
molecules, as it is the case for microgel suspensions in
water-trehalose solution (Fig. S2), the cumulant analysis does not
apply to $G^{(1)}(t)$, but only to its contribution due to the
Brownian motion of suspended microgel particles, as illustrated in
Fig. S3.

    \subsection{Supplementary figures}
In the following, we report additional information obtained from
experimental methods applied to PNIPAM microgel suspensions in water
and in water-trehalose solution at trehalose mole fraction
$x_{tr}=0.015$.

\begin{itemize}
  \item Figure S1 reports a characterization of the solvent
  media in which the microgels were suspended.
  \item Figure S2 shows the intensity autocorrelation function,
  $G^{(2)}(t)$, obtained by DLS at different temperatures for microgels
  in water-trehalose solution.
  \item Figure S3 illustrates the cumulants analysis performed on
  the DLS spectra of the microgel suspension in water-trehalose solution.
  \item Figure S4 compares the temperature dependence of the hydrodynamic radius,
  $R_h$, of microgels in water and in water-trehalose solution.
  \item Figure S5 illustrates the normalization procedure of the Raman
  spectra of microgel suspensions in water and in water-trehalose
  solution, needed to carry out the solvent subtraction procedure.
  \item Figure S6 illustrates the solvent subtraction procedure performed on
  the Raman spectra of microgel suspensions in water and in water-trehalose
  solution.
\end{itemize}

\newpage

\begin{figure}[h!]
\centering\includegraphics[width=1\linewidth]{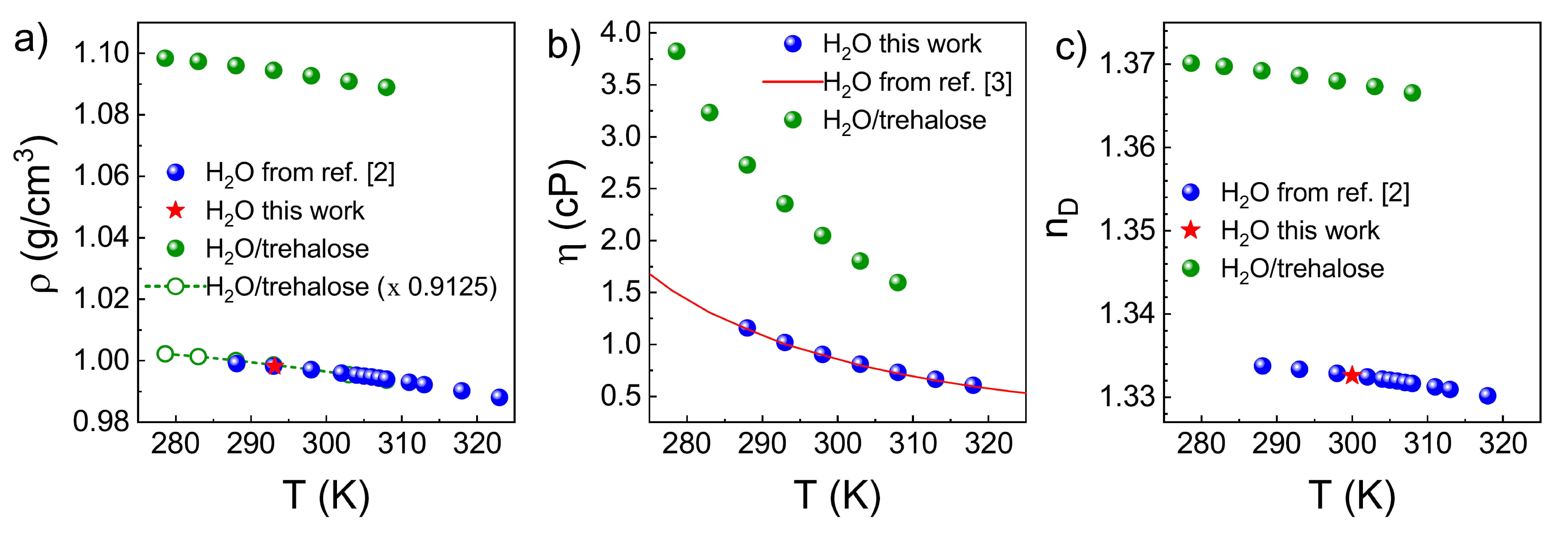} \caption{Fig.
S1: Characterization of the solvent media in which the microgels
were suspended. (a) Temperature dependence of mass density $\rho$,
(b) dynamic viscosity $\eta$, and (c) refractive index $n_D$, of
water and water-trehalose solution at trehalose mole fraction
$x_{tr}=0.015$. Panel (a) shows that the data of pure water, in the
range investigated, are scaled on those of water-trehalose solution
by a multiplication factor.} \label{fig:g2}
\end{figure}

\begin{figure}
\centering\includegraphics[width=0.9\linewidth]{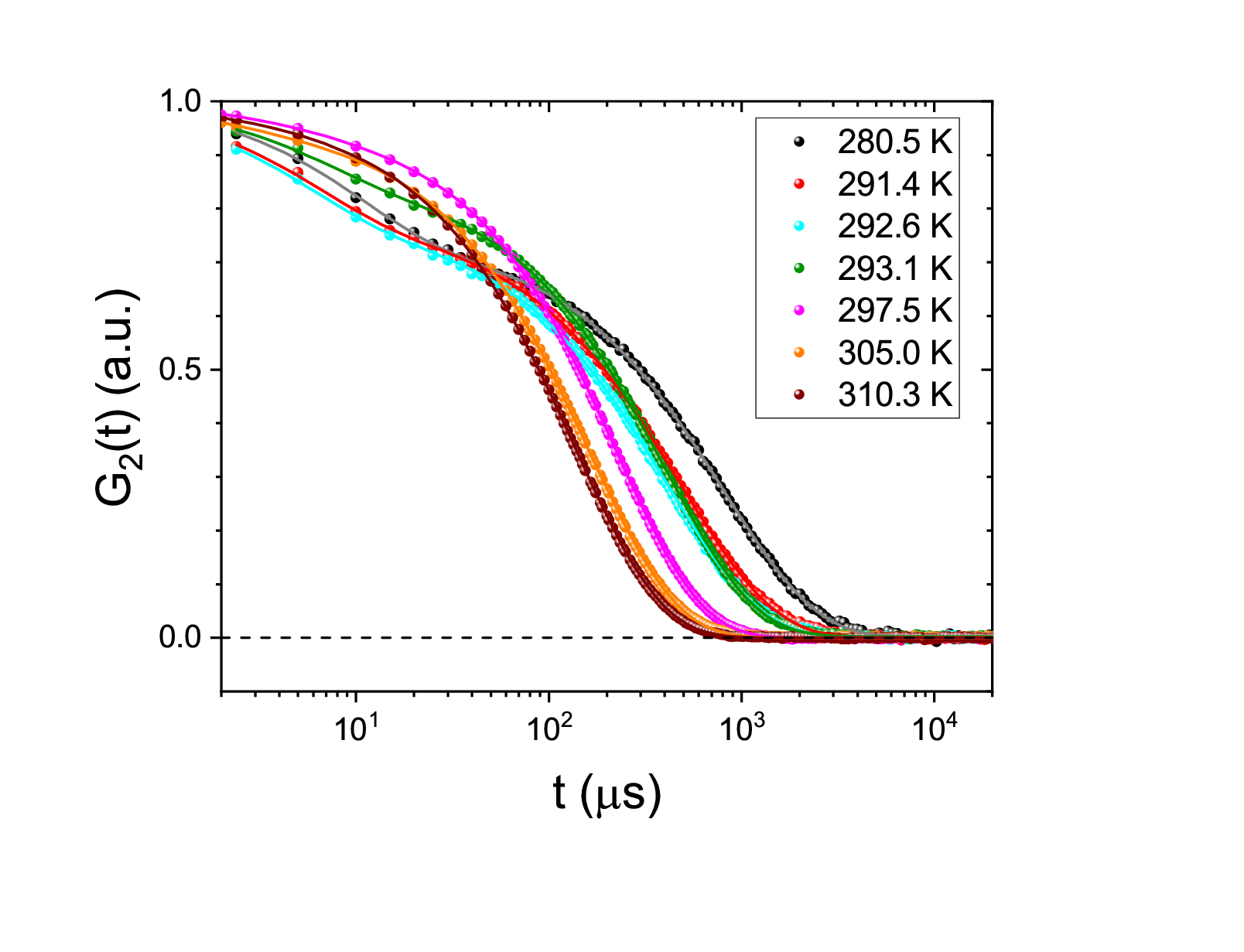} \caption{Fig.
S2: Intensity autocorrelation function, $G^{(2)}(t)=A_{0}\left[1 +
\beta |G^{(1)}(t)|^{2} \right]$, measured at different temperatures
in a highly diluted sample of PNIPAM microgels suspended in
water-trehalose solution. Symbols indicate experimental data, solid
lines represent the fitting curves obtained as explained in Fig. S3.
The correlation functions are normalized to their total amplitude as
obtained from the fit. The decay at small times reveals the presence
in the autocorrelation function of the scattered field,
$G^{(1)}(t)$, of a relaxation term due to the dynamics of trehalose
molecules in aqueous solution.} \label{fig:g2}
\end{figure}

\begin{figure}[h!]
\centering\includegraphics[width=1.1\linewidth]{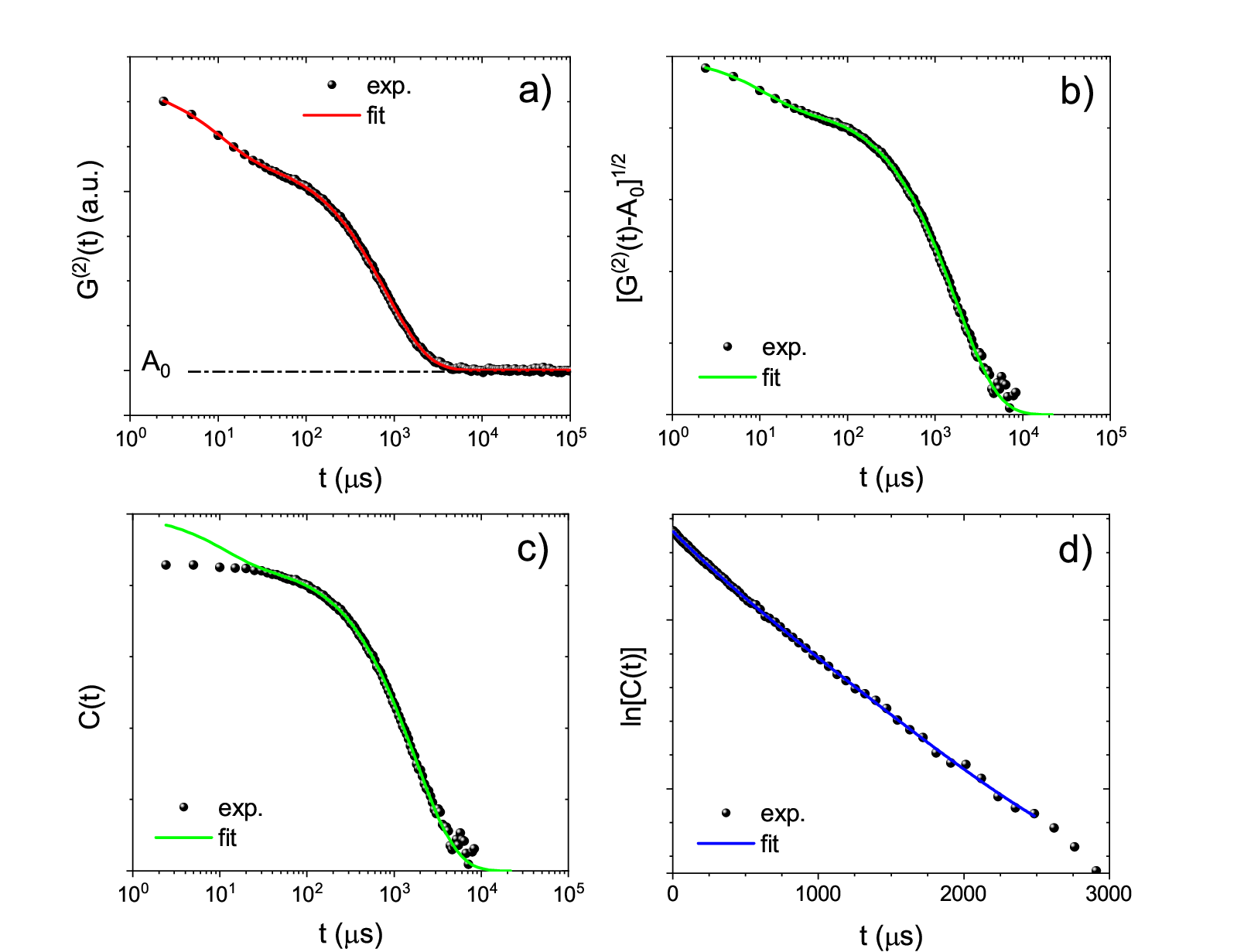}
\caption{Fig. S3: As an illustration, we show the steps followed to
analyze the DLS signal of the microgel suspension in water-trehalose
at 280.5 K. \textbf{First step:} The quantity
$G^{(2)}(t)=A_{0}\left[1 + \beta |G^{(1)}(t)|^{2} \right]$ directly
measured by DLS is shown in panel (a) with solid symbols.
\textbf{Second step:} $G^{(2)}(t)$ is used to calculate the quantity
$[G^{(2)}(t)-A_{0}]^{1/2}$, proportional to $G^{(1)}(t)$. This
quantity can be modeled as the superposition of two decay
contributions, the one at smaller times due to the relaxation of
trehalose in water, and the other due to the Brownian motion of
microgel particles in suspension. Both contributions are formally
reproduced by stretched exponential functions. Accordingly, the data
of panel (a) are fitted with $G^{(2)}(t)=A_{0}+\left[A_{1}\exp
(-(t/\tau_{1})^{\beta_{1}} + A_{2}\exp
(-(t/\tau_{2})^{\beta_{2}})\right]^{2}$, with $\beta_{1,2}<1$ the
stretching coefficients. Since the fit is rapidly convergent to
$\beta_{1}=1$ within the uncertainty, to minimize the number of free
fit-parameters the fitting procedure is performed at all
temperatures by setting $\beta_{1}=1$, i.e., by reproducing the
trehalose relaxation function with a simple exponential. The red
solid line in panel (a) demonstrates a perfect fit to the
experimental data. This line translates into the green line of panel
(b), also reported in panel (c) for comparison. \textbf{Third step:}
By subtracting from the data of panel (b) the quantity $A_{1}\exp
(-t/\tau_{1})$ obtained from the fit, the contribution $C(t)$ due to
the Brownian motion of microgel particles is finally obtained (panel
(c)). \textbf{Fourth step:} The deviation of $C(t)$ from a single
exponential decay is analyzed by the method of cumulants (panel
(d)). The logarithm of $C(t)$ is fitted by a fourth-order
polynomial, whose first and second-order terms are respectively
related to the z-average value ($<D>_{z}$) and variance ($<(\delta
D)^{2}>_{z}$) of the diffusion coefficient distribution of microgel
particles.} \label{fig:cum}
\end{figure}

\begin{figure}
\centering\includegraphics[width=0.8\linewidth]{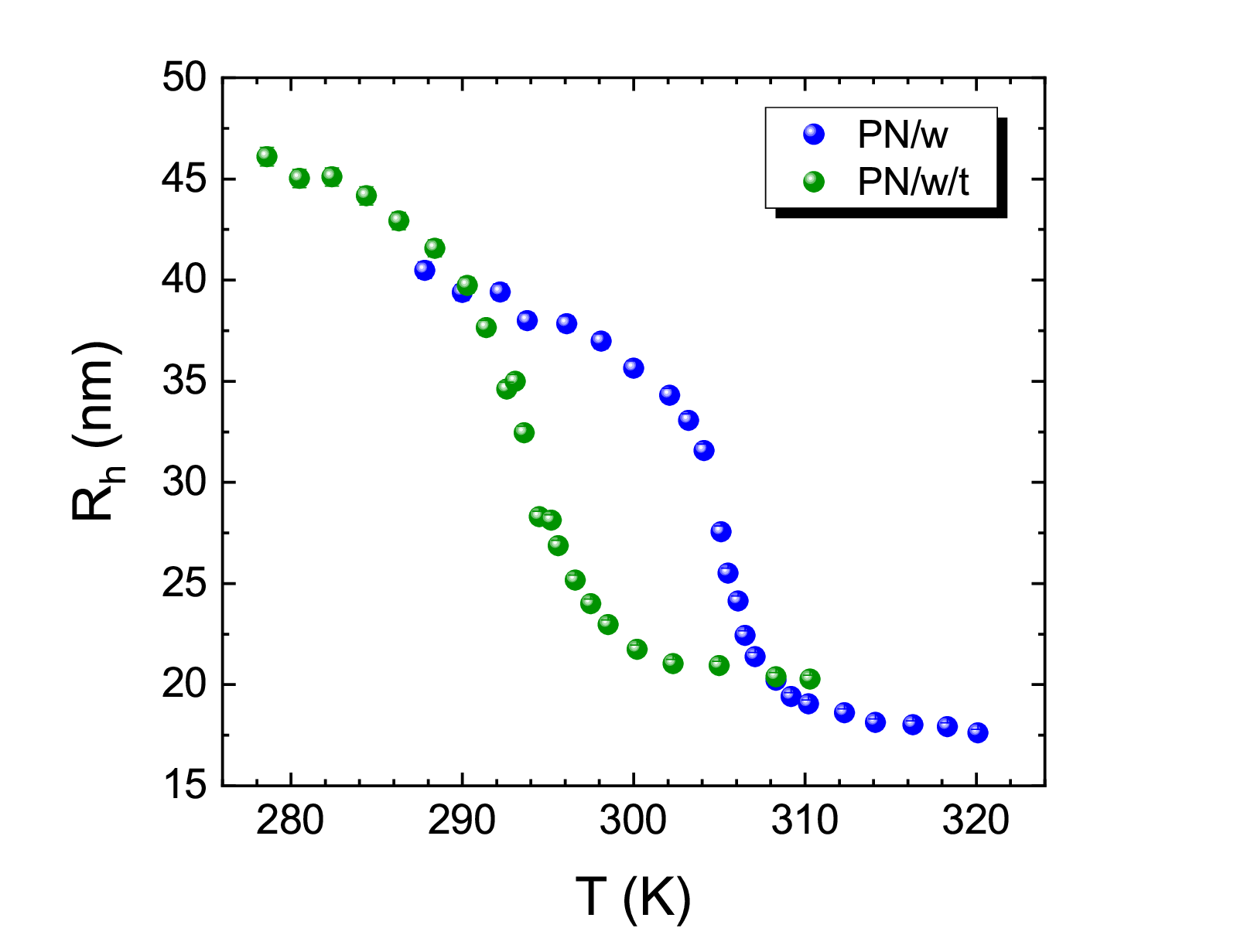}
\caption{Fig. S4: Hydrodynamic radius $R_h$ as a function of
temperature of PNIPAM microgels in water and water-trehalose. Errors
are within symbol size.} \label{fig:g2}
\end{figure}

\begin{figure}[h!]
\centering\includegraphics[width=0.8\linewidth]{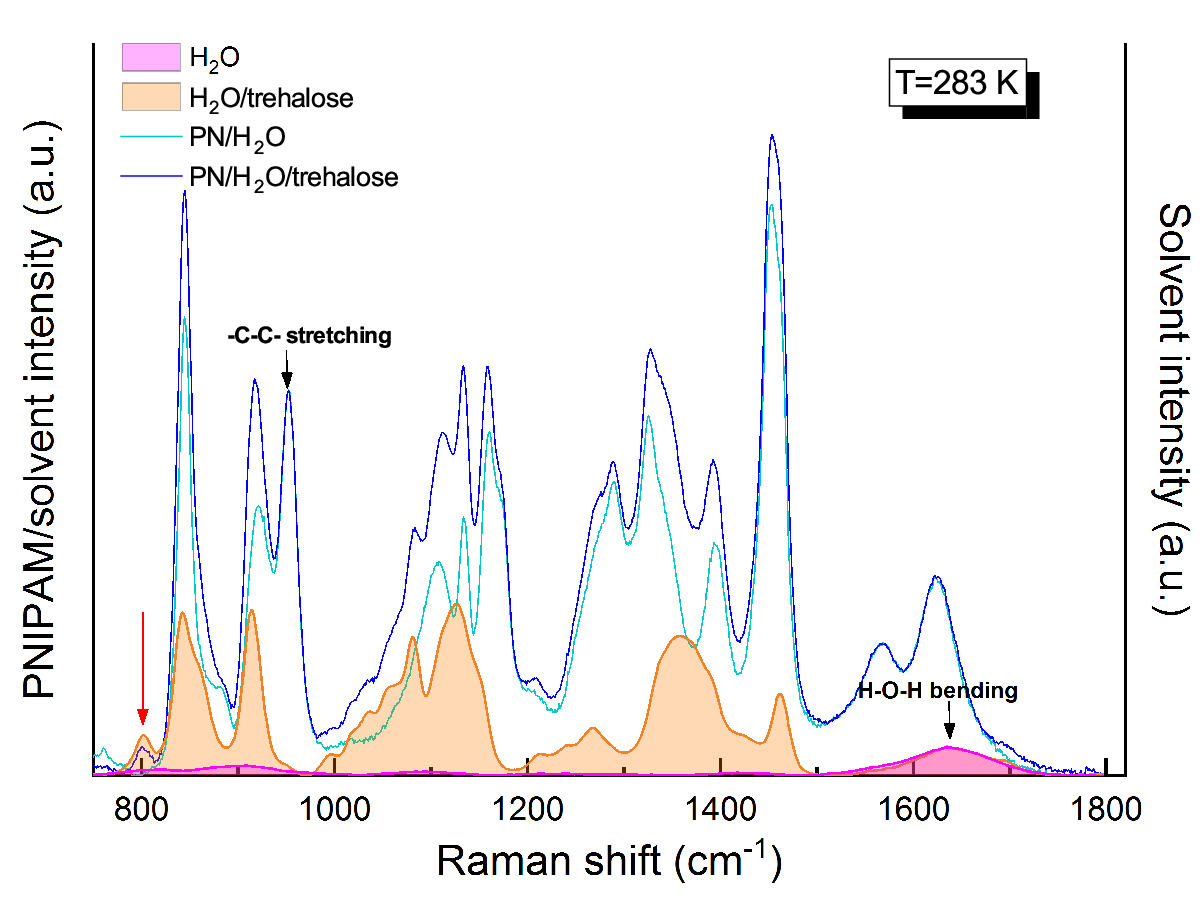}
\caption{Fig. S5: At each temperature, the spectra of PNIPAM
microgels in the two solvents, water and water-trehalose, were
normalized to the C-C stretching vibration peak at about 950
cm$^{-1}$ where both trehalose and water negligibly contribute to
the Raman signal, i.e., they were normalized to the sample's content
of PNIPAM. Analogously, the two solvent spectra were normalized to
the H-O-H bending mode of water in the 1500-1700 cm$^{-1}$ frequency
range, i.e., to the water content of the sample. The red arrow
indicates the peak at about 800 cm$^{-1}$, the only trehalose signal
with no superposition with those of PNIPAM. Notice that, for
graphical reasons, the solvent and PNIPAM/solvent spectra are
represented on different arbitrary scales.} \label{fig:Raman1}
\end{figure}

\begin{figure}[h!]
\centering\includegraphics[width=0.85\linewidth]{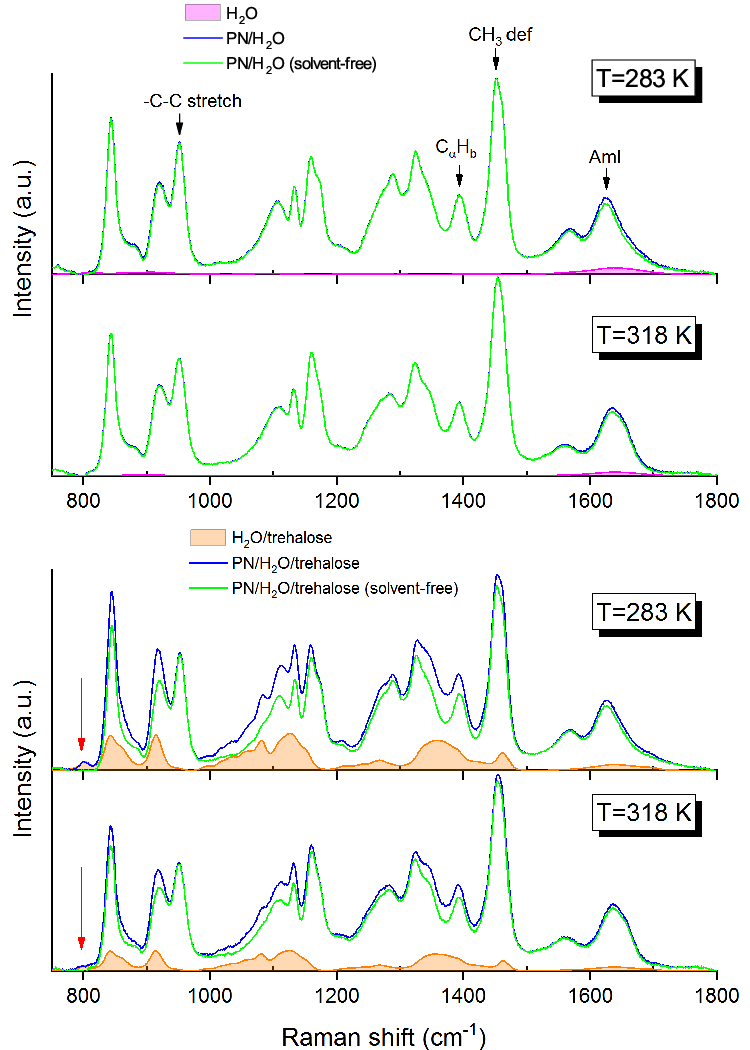}
\caption{Fig. S6: Starting from the normalized PNIPAM/solvent
spectra and the normalized solvent spectra as described in the
previous figure, the solvent-free spectra were obtained by
subtracting from each normalized PNIPAM/solvent spectrum the
corresponding normalized solvent signal, as much as to avoid
negative residues in the difference spectrum. It has to be noted
that in the trehalose-containing system this operation resulted in
the complete removal of the peak at about 800 cm$^{-1}$ (red arrow),
indicating that the subtraction was carried out at each temperature
with spectra put on the same relative scale, as shown in each panel
of the present figure. } \label{fig:Raman2}
\end{figure}

\clearpage

\section{Molecular dynamics simulations}

     \subsection{Model and simulation procedure}
The model of PNIPAM solution was assembled by centering the polymer
chain in a cubic box of side 9 nm. The chain conformation of the
starting structure was obtained by imposing values to the backbone
dihedral angles corresponding to states of minimum conformational
energy for the dyads composing the 30-mer \cite{FloryJACS1966}. The
chain was oriented along a diagonal of the cubic box to maximize the
distance between periodic images. Then, the proper number of water
or water and trehalose molecules was added randomly around the chain
to replicate the experimental solvent, i.e., pure water or
water-trehalose mixture at $x_{tr}=0.015$. Energy minimization was
carried out with tolerance of 100 kJ mol$^{-1}$nm$^{-1}$, and the
resulting configuration was used to start simulations at the two
studied temperatures (283 and 318 K). After equilibration the box
side was about 8.2 nm and 8.9 nm for the water-trehalose and the
aqueous solution, respectively. MD simulation trajectories were
acquired with the leapfrog integration algorithm
\cite{HockneyMCP1970} with a time step of 2 fs. Cubic periodic
boundary conditions and minimum image convention were applied. The
length of bonds involving hydrogen atoms was constrained with the
LINCS procedure \cite{HessJCC1997}. The temperature was controlled
by the velocity rescaling thermostat \cite{BussiJCP2007}, with a
time constant of 0.1 ps. The pressure of 1 atm was maintained by the
Parrinello-Rahman barostat \cite{ParrinelloJAP1981, NoseMP1983},
with a time constant of 2 ps. Electrostatic interactions were
treated with the smooth particle-mesh Ewald method
\cite{EssmannJCP1995} with a cutoff of nonbonded interactions of 1
nm.

     \subsection{Supplementary figures}
In the following, we report additional characterizations of the
solution behavior of PNIPAM in water and in water-trehalose at
trehalose mole fraction $x_{tr}=0.015$, as obtained from the MD
simulations.

\begin{itemize}
  \item Figure S7 compares the distribution of values of PNIPAM radius
  of gyration, $R_g$, in water and in water-trehalose.
  \item Figure S8 compares the distribution of values of water accessible
  surface area (WASA) of PNIPAM in water and in water-trehalose.
  \item Figure S9 shows the radial distribution functions, $g(r)$,
  calculated between the anomeric oxygen of trehalose and specific atoms of
  PNIPAM belonging to hydrophilic and hydrophobic groups.
  \item Figure S10 reports the radial distribution functions characterizing
  PNIPAM hydration.
  \item Figure S11 shows the water coordination number, $CN(r)$, of PNIPAM
  specific groups, as obtained by integration of the corresponding radial
  distribution functions reported in Fig. S10.
  \item Figure S12 shows a comparison between the mean square displacements
  (MSDs) of water oxygen atoms, PNIPAM hydrogen atoms, and trehalose hydrogen atoms.
  \item Figure S13 reports the time behavior of the number fraction of
  trehalose and water molecules in the first solvation shell.
\end{itemize}

\begin{figure}[h!]
\centering\includegraphics[width=0.6\linewidth]{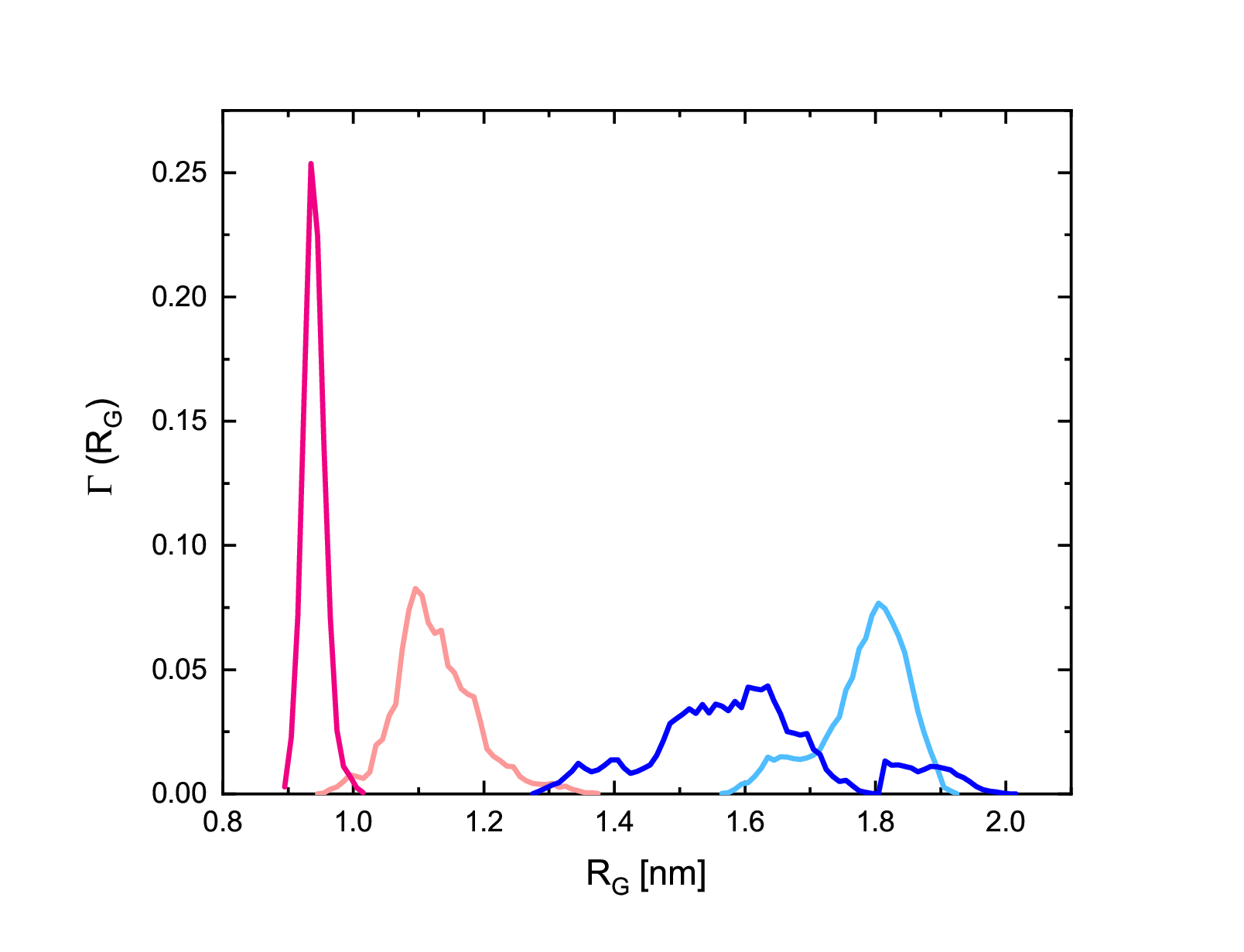}
\caption{Fig. S7: Distribution of values of PNIPAM $R_g$ in water
(blue and magenta line at 283 K and 318 K, respectively) and in
water-trehalose (light blue and pink line at 283 K and 318 K,
respectively).} \label{fig:Rg_distr}
\end{figure}

\begin{figure}[h!]
\centering\includegraphics[width=0.6\linewidth]{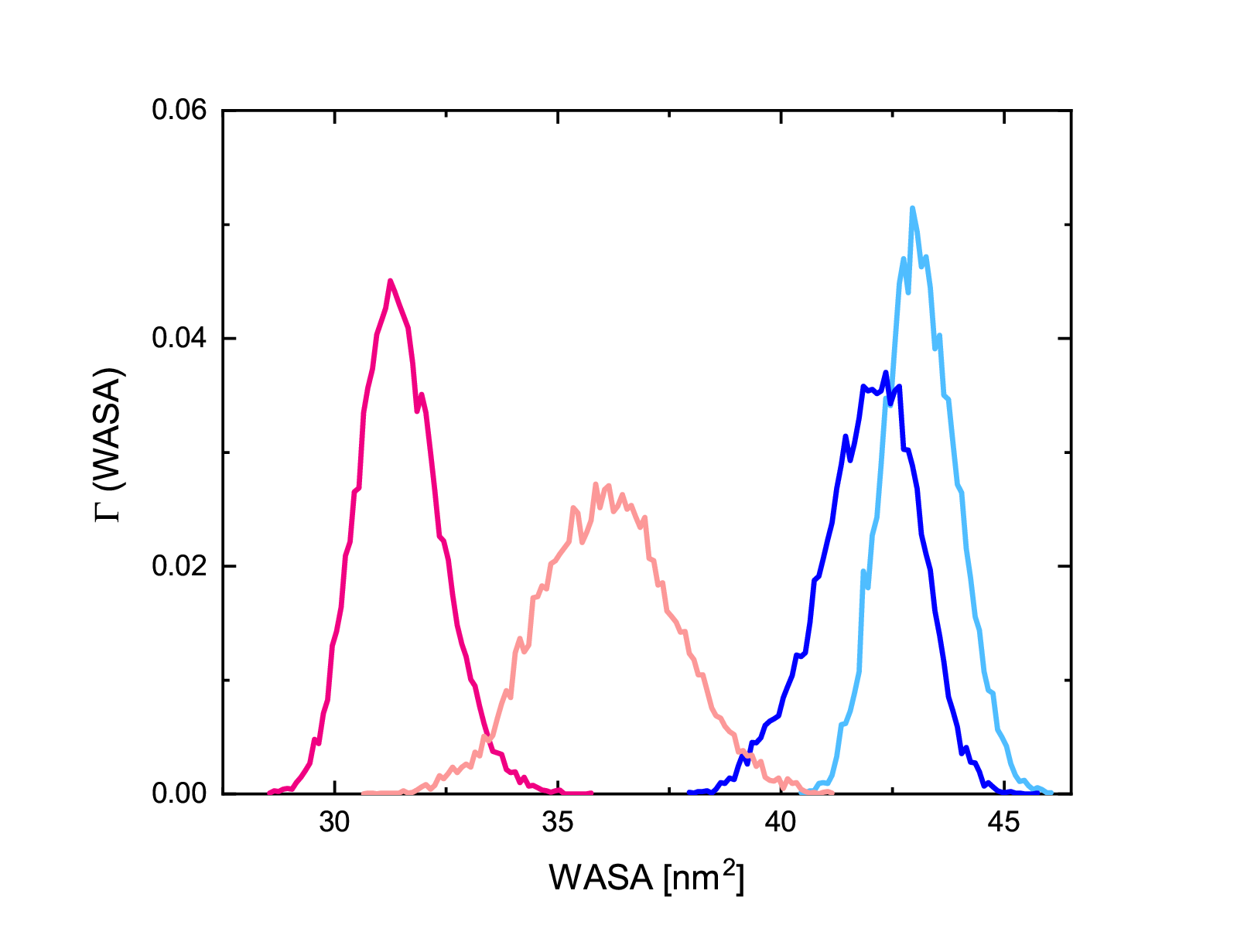} \caption{Fig.
S8: Distribution of values of WASA for PNIPAM in water (blue and
magenta line at 283 K and 318 K, respectively) and in
water-trehalose (light blue and pink line at 283 K and 318 K,
respectively).} \label{fig:WASA2}
\end{figure}

\begin{figure}[h!]
\centering\includegraphics[width=1\linewidth]{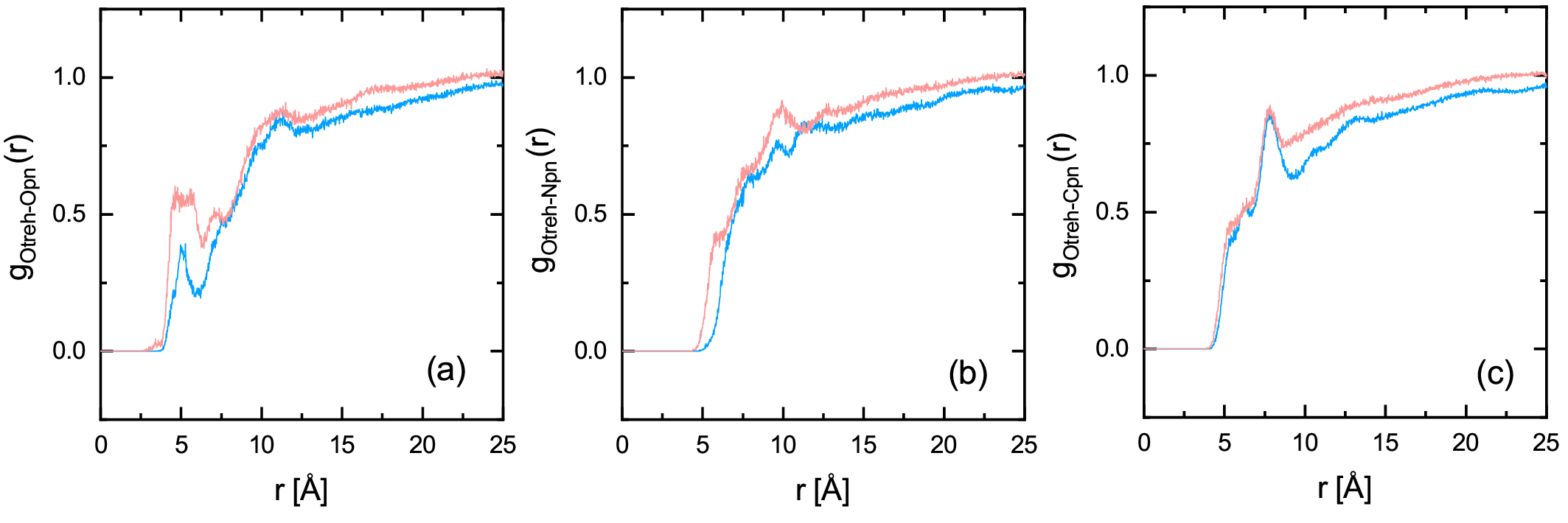}
\caption{Fig. S9: Radial distribution function g(r) between the
anomeric oxygen atom of trehalose and PNIPAM (a) oxygen, (b)
nitrogen and (c) methyl carbon atoms, calculated at 283 K (blue
line) and 318 K (pink line).} \label{fig:rdfO1}
\end{figure}

\begin{figure}[h!]
\centering\includegraphics[width=0.9\linewidth]{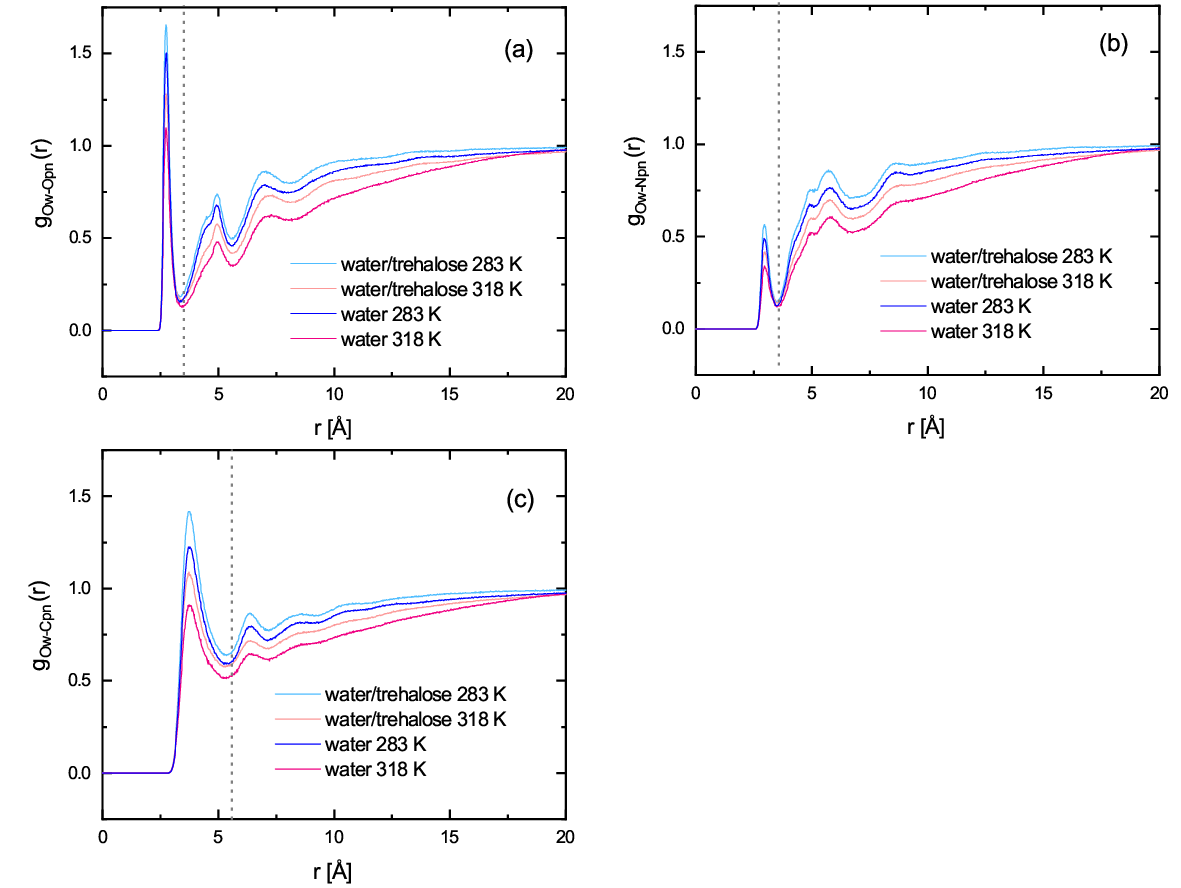}
\caption{Fig. S10: Radial distribution function g(r) between water
oxygen and PNIPAM (a) oxygen, (b) nitrogen, and (c) methyl carbon
atoms, calculated at 283 and 318 K. Vertical dashed lines indicate
the thickness of the first solvation shell. The higher value of the
radial distribution functions of the water-trehalose-PNIPAM system
are due to a normalization effect.} \label{fig:rdf}
\end{figure}

\begin{figure}[h!]
\centering\includegraphics[width=0.85\linewidth]{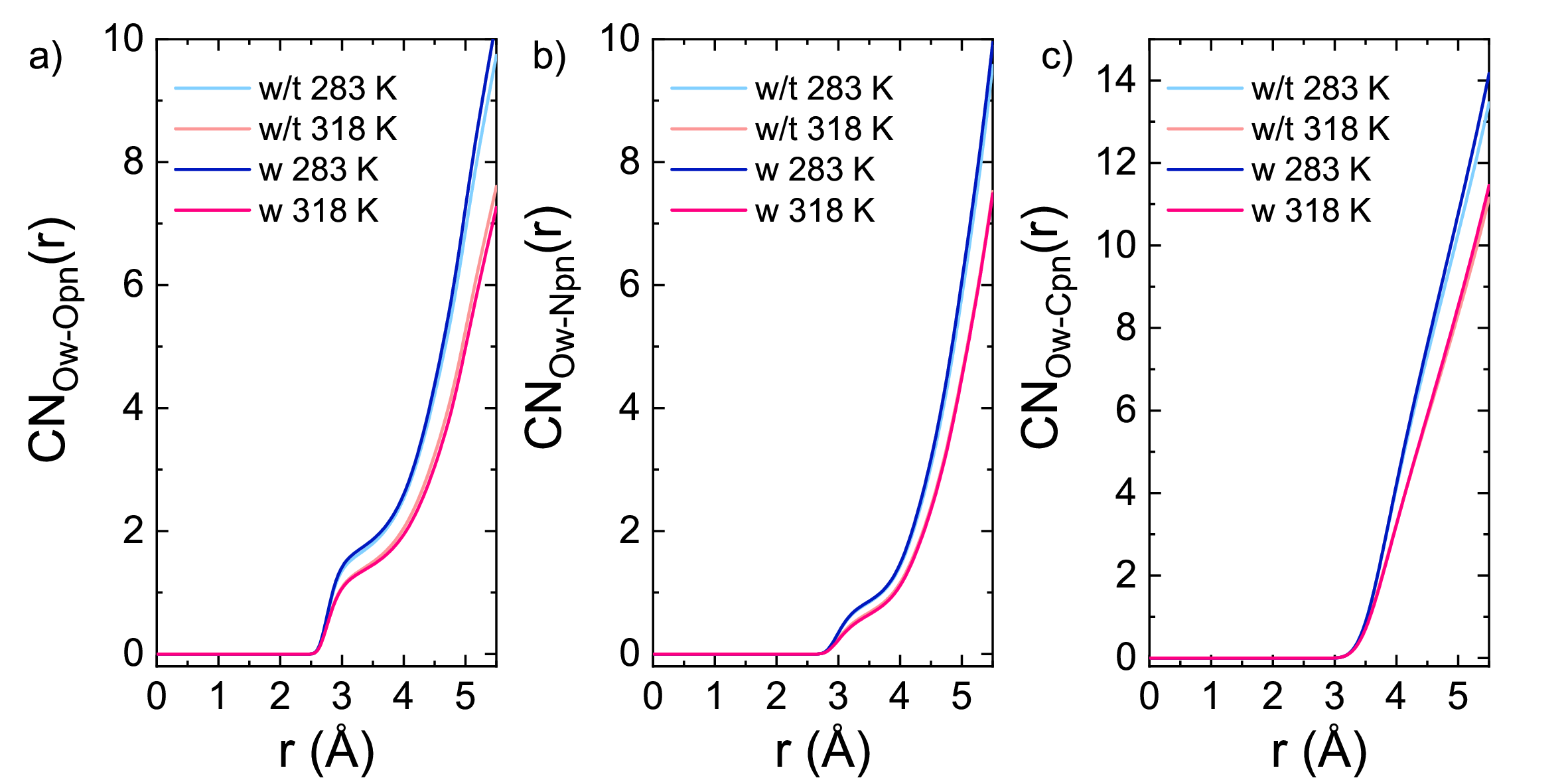} \caption{Fig.
S11: Coordination number CN(r) of water oxygen atoms around PNIPAM
(a) oxygen, (b) nitrogen and (c) methyl carbon atoms, calculated at
283 and 318 K.} \label{fig:CN}
\end{figure}

\begin{figure}[h!]
\centering\includegraphics[width=1\linewidth]{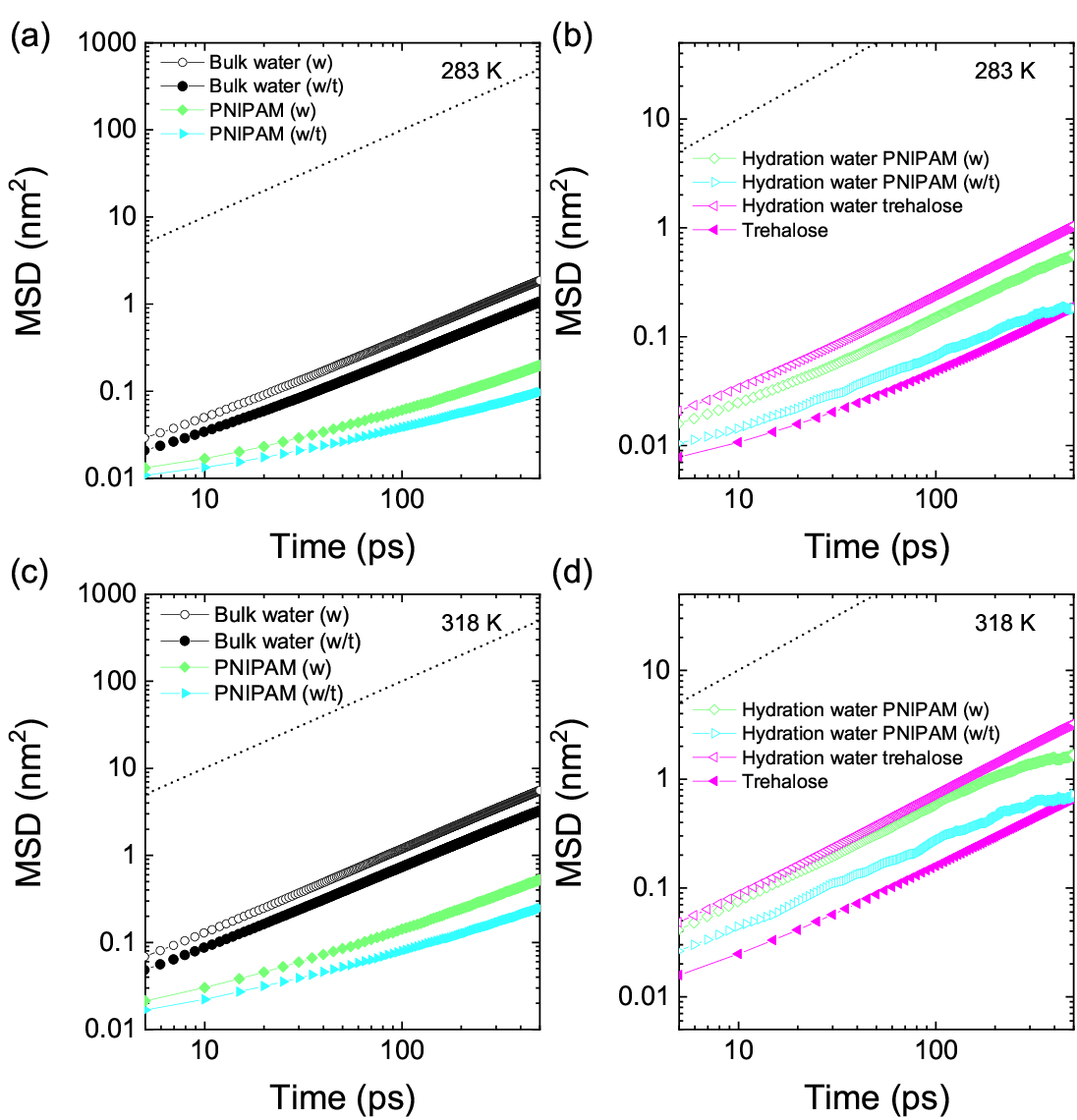} \caption{Fig.
S12: Mean square displacements (MSDs) of oxygen atoms of bulk and
hydration water and of hydrogen atoms of PNIPAM and trehalose,
calculated at 283 K (upper panels) and 318 K (lower panels) as
indicated in the legend. In all panels a dotted black line with
slope 1 is displayed.} \label{fig:msd}
\end{figure}

\clearpage

\begin{figure}[h!]
\centering\includegraphics[width=1\linewidth]{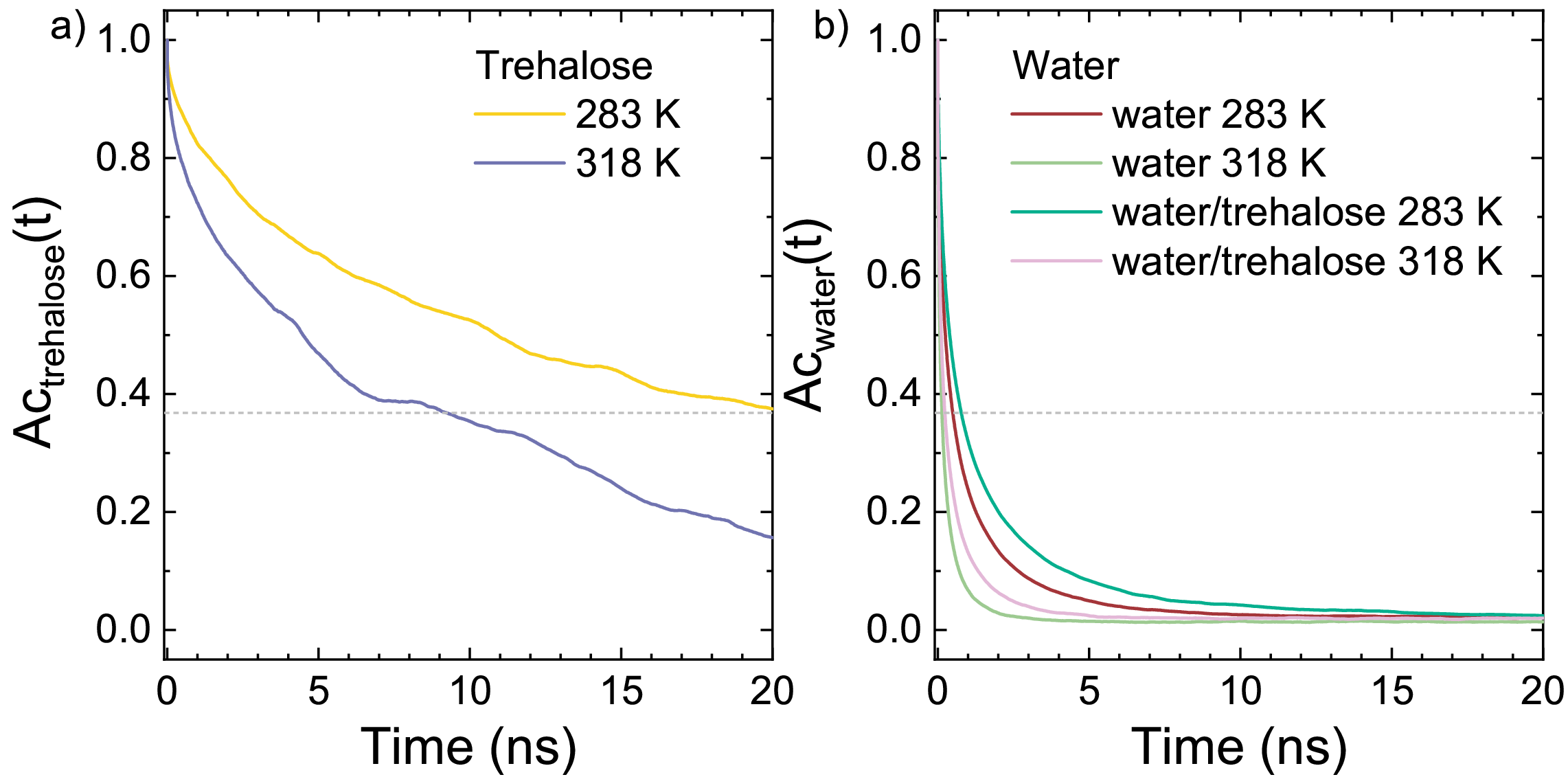} \caption{Fig.
S13: Autocorrelation function of the number fraction of trehalose
(a) and water (b) molecules residing in the first solvation shell,
calculated at 283 and 318 K.} \label{fig:Restime}
\end{figure}

\section*{Supplementary references}

{}


\section*{References}

\begin{thebibliography}{82}
\expandafter\ifx\csname url\endcsname\relax
  \def\url#1{\texttt{#1}}\fi
\expandafter\ifx\csname urlprefix\endcsname\relax\def\urlprefix{URL }\fi
\expandafter\ifx\csname href\endcsname\relax
  \def\href#1#2{#2} \def\path#1{#1}\fi

\bibitem{BookProteinPhysics}
A.~Finkelstein, O.~Ptitsyn, Protein Physics, Academic Press, Elsevier, 2016.

\bibitem{BookProteinInteractions}
J.~W. Shriver (Ed.), Protein Structure, Stability, and Interactions, Humana
  Press, 2009.
\newblock \href {http://dx.doi.org/10.1007/978-1-59745-367-7}
  {\path{doi:10.1007/978-1-59745-367-7}}.

\bibitem{CostaPOLYMER2002}
R.~O. Costa, R.~F. Freitas, Phase behavior of poly (n-isopropylacrylamide) in
  binary aqueous solutions, Polymer 43~(22) (2002) 5879--5885.
\newblock \href {http://dx.doi.org/10.1016/S0032-3861(02)00507-4}
  {\path{doi:10.1016/S0032-3861(02)00507-4}}.

\bibitem{YamauchiJPCB2007}
H.~Yamauchi, Y.~Maeda, Lcst and ucst behavior of poly(n-isopropylacrylamide) in
  dmso/water mixed solvents studied by ir and micro-raman spectroscopy, The
  Journal of Physical Chemistry B 111~(45) (2007) 12964--12968.
\newblock \href {http://dx.doi.org/10.1021/jp072438s}
  {\path{doi:10.1021/jp072438s}}.

\bibitem{PerezAPM2019}
H.~P{\'e}rez-Ram{\'\i}rez, C.~Haro-P{\'e}rez, G.~Odriozola, Effect of
  temperature on the cononsolvency of poly (n-isopropylacrylamide)(pnipam) in
  aqueous 1-propanol, ACS Applied Polymer Materials 1~(11) (2019) 2961--2972.
\newblock \href {http://dx.doi.org/10.1021/acsapm.9b00665}
  {\path{doi:10.1021/acsapm.9b00665}}.

\bibitem{ZhangJACS2005}
Y.~Zhang, S.~Furyk, D.~E. Bergbreiter, P.~S. Cremer, Specific ion effects on
  the water solubility of macromolecules: Pnipam and the hofmeister series,
  Journal of the American Chemical Society 127~(41) (2005) 14505--14510.
\newblock \href {http://dx.doi.org/10.1021/ja0546424}
  {\path{doi:10.1021/ja0546424}}.

\bibitem{HeydaJPCB2014}
J.~Heyda, J.~Dzubiella, Thermodynamic description of hofmeister effects on the
  lcst of thermosensitive polymers, The Journal of Physical Chemistry B
  118~(37) (2014) 10979--10988.
\newblock \href {http://dx.doi.org/10.1021/jp5041635}
  {\path{doi:10.1021/jp5041635}}.

\bibitem{PicaPCCP2020}
A.~Pica, G.~Graziano, Effect of sodium thiocyanate and sodium perchlorate on
  poly(n-isopropylacrylamide) collapse, Physical Chemistry Chemical Physics
  22~(1) (2020) 189--195.
\newblock \href {http://dx.doi.org/10.1039/C9CP05706D}
  {\path{doi:10.1039/C9CP05706D}}.

\bibitem{ShpigelmanJPSB2008}
A.~Shpigelman, I.~Portnaya, O.~Ramon, Y.~D. Livney, Saccharide-structure
  effects on poly $n$-isopropylacrylamide phase transition in aqueous media;
  reflections on protein stability, Journal of Polymer Science Part B: Polymer
  Physics 46 (2008) 2307--2318.
\newblock \href {http://dx.doi.org/10.1002/polb.21562}
  {\path{doi:10.1002/polb.21562}}.

\bibitem{SagleJACS2009}
L.~B. Sagle, Y.~Zhang, V.~A. Litosh, X.~Chen, Y.~Cho, P.~S. Cremer,
  Investigating the hydrogen-bonding model of urea denaturation, Journal of the
  American Chemical Society 131~(26) (2009) 9304--9310.
\newblock \href {http://dx.doi.org/10.1021/ja9016057}
  {\path{doi:10.1021/ja9016057}}.

\bibitem{KimEPJ2014}
S.~M. Kim, S.~M. Lee, Y.~C. Bae, Influence of hydroxyl group for
  thermoresponsive poly(n-isopropylacrylamide) gel particles in
  water/co-solvent (1,3-propanediol, glycerol) systems, European Polymer
  Journal 54 (2014) 151--159.
\newblock \href {http://dx.doi.org/10.1016/j.eurpolymj.2014.03.008}
  {\path{doi:10.1016/j.eurpolymj.2014.03.008}}.

\bibitem{SchroerPCCP2016}
M.~A. Schroer, J.~Michalowsky, B.~Fischer, J.~Smiatek, G.~Grübel, Stabilizing
  effect of tmao on globular pnipam states: preferential attraction induces
  preferential hydration, Phys. Chem. Chem. Phys. 18~(46) (2016) 31459--31470.
\newblock \href {http://dx.doi.org/10.1039/C6CP05991K}
  {\path{doi:10.1039/C6CP05991K}}.

\bibitem{NarangNJC2018}
P.~Narang, P.~Venkatesu, Unravelling the role of polyols with increasing carbon
  chain length and oh groups on the phase transition behavior of pnipam, New J.
  Chem. 42~(16) (2018) 13708--13717.
\newblock \href {http://dx.doi.org/10.1039/C8NJ02510J}
  {\path{doi:10.1039/C8NJ02510J}}.

\bibitem{UmapathiJML2020}
R.~Umapathi, I.~Khan, J.~A. Coutinho, P.~Venkatesu, Unravelling the
  interactions between biomedical thermoresponsive polymer and biocompatible
  ionic liquids, Journal of Molecular Liquids 300 (2020) 112362.
\newblock \href {http://dx.doi.org/10.1016/j.molliq.2019.112362}
  {\path{doi:10.1016/j.molliq.2019.112362}}.

\bibitem{UmapathiJCIS2019}
R.~Umapathi, K.~Kumar, G.~M. Rani, P.~Venkatesu, Influence of biological
  stimuli on the phase behaviour of a biomedical thermoresponsive polymer: A
  comparative investigation of hemeproteins, Journal of Colloid and Interface
  Science 541 (2019) 1--11.
\newblock \href {http://dx.doi.org/10.1016/j.jcis.2019.01.062}
  {\path{doi:10.1016/j.jcis.2019.01.062}}.

\bibitem{NarangACIS2019}
P.~Narang, P.~Venkatesu, Efficacy of several additives to modulate the phase
  behavior of biomedical polymers: A comprehensive and comparative outlook,
  Advances in Colloid and Interface Science 274 (2019) 102042.
\newblock \href {http://dx.doi.org/10.1016/j.cis.2019.102042}
  {\path{doi:10.1016/j.cis.2019.102042}}.

\bibitem{MartinezMoroJCIS2020}
M.~Martinez-Moro, J.~Jenczyk, J.~M. Giussi, S.~Jurga, S.~E. Moya, Kinetics of
  the thermal response of poly(n-isopropylacrylamide co methacrylic acid)
  hydrogel microparticles under different environmental stimuli: A time-lapse
  nmr study, Journal of Colloid and Interface Science 580 (2020) 439--448.
\newblock \href {http://dx.doi.org/10.1016/j.jcis.2020.07.049}
  {\path{doi:10.1016/j.jcis.2020.07.049}}.

\bibitem{KumarLANG2021}
K.~Kumar, R.~Umapathi, K.~Ramesh, S.-K. Hwang, K.~T. Lim, Y.~S. Huh,
  P.~Venkatesu, Biological stimuli-induced phase transition of a synthesized
  block copolymer: Preferential interactions between pnipam-b-pnvcl and heme
  proteins, Langmuir 37~(5) (2021) 1682--1696.
\newblock \href {http://dx.doi.org/10.1021/acs.langmuir.0c02900}
  {\path{doi:10.1021/acs.langmuir.0c02900}}.

\bibitem{TiktopuloMACRO1994}
E.~I. Tiktopulo, V.~E. Bychkova, J.~Ri\v{c}ka, O.~B. Ptitsyn, Cooperativity of
  the coil-globule transition in a homopolymer: microcalorimetric study of
  poly($n$-isopropylacrylamide), Macromolecules 27 (1994) 2879--2882.
\newblock \href {http://dx.doi.org/10.1021/ma00088a031}
  {\path{doi:10.1021/ma00088a031}}.

\bibitem{ZanattaSCIADV2018}
M.~Zanatta, L.~Tavagnacco, E.~Buratti, M.~Bertoldo, F.~Natali, E.~Chiessi,
  A.~Orecchini, E.~Zaccarelli, Evidence of a low-temperature dynamical
  transition in concentrated microgels, Science Advances 4~(9).
\newblock \href {http://dx.doi.org/10.1126/sciadv.aat5895}
  {\path{doi:10.1126/sciadv.aat5895}}.

\bibitem{BookFernandez2011}
A.~Fernandez-Nieves, H.~Wyss, J.~Mattsson, D.~Weitz (Eds.), Microgel
  suspensions: fundamentals and applications, Wiley-VCH Verlag, Germany, 2011.
\newblock \href {http://dx.doi.org/10.1002/9783527632992}
  {\path{doi:10.1002/9783527632992}}.

\bibitem{BeattieDIAB1997}
G.~M. Beattie, J.~Crowe, A.~Lopez, V.~Cirulli, C.~Ricordi, A.~Hayek, Trehalose:
  a cryoprotectant that enhances recovery and preserves function of human
  pancreatic islets after long-term storage, Diabetes 46~(3) (1997) 519--523.
\newblock \href {http://dx.doi.org/10.2337/diab.46.3.519}
  {\path{doi:10.2337/diab.46.3.519}}.

\bibitem{CarninciPNAS1998}
P.~Carninci, Y.~Nishiyama, A.~Westover, M.~Itoh, S.~Nagaoka, N.~Sasaki,
  Y.~Okazaki, M.~Muramatsu, Y.~Hayashizaki, Thermostabilization and
  thermoactivation of thermolabile enzymes by trehalose and its application for
  the synthesis of full length cdna, Proceedings of the National Academy of
  Sciences 95~(2) (1998) 520--524.
\newblock \href {http://dx.doi.org/10.1073/pnas.95.2.520}
  {\path{doi:10.1073/pnas.95.2.520}}.

\bibitem{BenaroudjJBC2001}
N.~Benaroudj, D.~Lee, A.~Goldberg, Trehalose accumulation during cellular
  stress protects cells and cellular proteins from damage by oxygen radicals,
  The Journal of Biological Chemistry 276~(26) (2001) 24261--24267.
\newblock \href {http://dx.doi.org/10.1074/jbc.m101487200}
  {\path{doi:10.1074/jbc.m101487200}}.

\bibitem{CesaroNATMAT2006}
A.~Ces{\`a}ro, All dried up, Nature materials 5~(8) (2006) 593--594.

\bibitem{KaushikJBC2003}
J.~K. Kaushik, R.~Bhat, Why is trehalose an exceptional protein stabilizer? an
  analysis of the thermal stability of proteins in the presence of the
  compatible osmolyte trehalose, Journal of Biological Chemistry 278~(29)
  (2003) 26458--26465.
\newblock \href {http://dx.doi.org/10.1074/jbc.M300815200}
  {\path{doi:10.1074/jbc.M300815200}}.

\bibitem{JainPROTSCI2009}
N.~K. Jain, I.~Roy, Effect of trehalose on protein structure, Protein Science
  18~(1) (2009) 24--36.
\newblock \href {http://dx.doi.org/https://doi.org/10.1002/pro.3}
  {\path{doi:https://doi.org/10.1002/pro.3}}.

\bibitem{TimasheffBIOCHEM1982}
T.~Arakawa, S.~N. Timasheff, Stabilization of protein structure by sugars,
  Biochemistry 21 (1982) 6536--6544.
\newblock \href {http://dx.doi.org/10.1021/bi00268a033}
  {\path{doi:10.1021/bi00268a033}}.

\bibitem{TimasheffBIOCHEM1997}
G.~Xie, S.~N. Timasheff, The thermodynamic mechanism of protein stabilization
  by trehalose, Biophysical Chemistry 64~(1) (1997) 25--43.
\newblock \href {http://dx.doi.org/10.1016/S0301-4622(96)02222-3}
  {\path{doi:10.1016/S0301-4622(96)02222-3}}.

\bibitem{CiceroneBIOPHYSJ2004}
M.~T. Cicerone, C.~L. Soles, Fast dynamics and stabilization of proteins:
  binary glasses of trehalose and glycerol, Biophysical Journal 86 (2004)
  3836--3845.
\newblock \href {http://dx.doi.org/10.1529/biophysj.103.035519}
  {\path{doi:10.1529/biophysj.103.035519}}.

\bibitem{SubatraJPCB2015}
S.~Paul, S.~Paul, Molecular insights into the role of aqueous trehalose
  solution on temperature-induced protein denaturation, The Journal of Physical
  Chemistry B 119~(4) (2015) 1598--1610.
\newblock \href {http://dx.doi.org/10.1021/jp510423n}
  {\path{doi:10.1021/jp510423n}}.

\bibitem{MalferrariJPCL2016}
M.~Malferrari, A.~Savitsky, W.~Lubitz, K.~Möbius, G.~Venturoli, Protein
  immobilization capabilities of sucrose and trehalose glasses: The effect of
  protein/sugar concentration unraveled by high-field epr, The Journal of
  Physical Chemistry Letters 7~(23) (2016) 4871--4877.
\newblock \href {http://dx.doi.org/10.1021/acs.jpclett.6b02449}
  {\path{doi:10.1021/acs.jpclett.6b02449}}.

\bibitem{CorradiniSCIREP2013}
D.~Corradini, E.~G. Strekalova, H.~E. Stanley, P.~Gallo, Microscopic mechanism
  of protein cryopreservation in an aqueous solution with trehalose, Scientific
  reports 3 (2013) 1218.
\newblock \href {http://dx.doi.org/10.1038/srep01218}
  {\path{doi:10.1038/srep01218}}.

\bibitem{CorezziJCP2019}
S.~Corezzi, M.~Paolantoni, P.~Sassi, A.~Morresi, D.~Fioretto, L.~Comez,
  Trehalose-induced slowdown of lysozyme hydration dynamics probed by
  \textsc{EDLS} spectroscopy, Journal of Chemical Physics 151.

\bibitem{CamisascaJCP2020}
G.~Camisasca, M.~De~Marzio, P.~Gallo, Effect of trehalose on protein
  cryoprotection: Insights into the mechanism of slowing down of hydration
  water, The Journal of Chemical Physics 153~(22) (2020) 224503.
\newblock \href {http://dx.doi.org/10.1063/5.0033526}
  {\path{doi:10.1063/5.0033526}}.

\bibitem{NarangJCIS2017}
P.~Narang, S.~B. Vepuri, P.~Venkatesu, M.~E. Soliman, An unexplored remarkable
  pnipam-osmolyte interaction study: An integrated experimental and simulation
  approach, Journal of Colloid and Interface Science 504 (2017) 417--428.
\newblock \href {http://dx.doi.org/10.1016/j.jcis.2017.05.109}
  {\path{doi:10.1016/j.jcis.2017.05.109}}.

\bibitem{Karg2013}
M.~Karg, S.~Pr\'{e}vost, A.~Brandt, D.~Wallacher, R.~von Klitzing, T.~Hellweg,
  Poly-NIPAM microgels with different cross-linker densities. In: Intelligent
  Hydrogels. Progress in Colloid and Polymer Science, Springer, Berlin, 2013.

\bibitem{ArlethJPSB2005}
L.~Arleth, X.~Xia, R.~Hjelm, J.~Wu, Z.~Hu, Volume transition and internal
  structures of small poly(n-isopropylacrylamide) microgels, Journal of Polymer
  Science. Part B, Polymer Physics 43 (2005) 849--860.
\newblock \href {http://dx.doi.org/10.1002/polb.20375}
  {\path{doi:10.1002/polb.20375}}.

\bibitem{AnderssonJPSB2006}
M.~Andersson, S.~L. Maunu, Structural studies of poly(n-isopropylacrylamide)
  microgels: Effect of sds surfactant concentration in the microgel synthesis,
  Journal of Polymer Science Part B: Polymer Physics 44~(23) (2006) 3305--3314.
\newblock \href {http://dx.doi.org/10.1002/polb.20971}
  {\path{doi:10.1002/polb.20971}}.

\bibitem{PhilippSM2012}
M.~Philipp, U.~Muller, R.~Aleksandrova, R.~Sanctuary, P.~Muller-Buschbaum,
  J.~K. Kruger, On the elastic nature of the demixing transition of aqueous
  pnipam solutions, Soft Matter 8 (2012) 11387--11395.
\newblock \href {http://dx.doi.org/10.1039/C2SM26527C}
  {\path{doi:10.1039/C2SM26527C}}.

\bibitem{KorsonJPC1968}
L.~Korson, W.~Drost-Hansen, F.~J. Millero, Viscosity of water at various
  temperatures, The Journal of Physical Chemistry 73~(1) (1969) 34--39.
\newblock \href {http://dx.doi.org/10.1021/j100721a006}
  {\path{doi:10.1021/j100721a006}}.

\bibitem{JorgensenJACS1996}
W.~L. Jorgensen, D.~S. Maxwell, J.~Tirado-Rives, Development and testing of the
  opls all-atom force field on conformational energetics and properties of
  organic liquids, J. Am. Chem. Soc. 118~(45) (1996) 11225--11236.
\newblock \href {http://dx.doi.org/10.1021/ja9621760}
  {\path{doi:10.1021/ja9621760}}.

\bibitem{SiuJCTC2012}
S.~W.~I. Siu, K.~Pluhackova, R.~A. B{\"o}ckmann, Optimization of the opls-aa
  force field for long hydrocarbons, J. Chem. Theory Comput. 8~(4) (2012)
  1459--1470.
\newblock \href {http://dx.doi.org/10.1021/ct200908r}
  {\path{doi:10.1021/ct200908r}}.

\bibitem{DammJCC1997}
W.~Damm, A.~Frontera, J.~Tirado-Rives, W.~L. Jorgensen, Opls all-atom force
  field for carbohydrates, Journal of Computational Chemistry 18~(16) (1997)
  1955--1970.

\bibitem{AbascalJCP2005}
J.~L.~F. Abascal, E.~Sanz, R.~G. Fern{\'a}ndez, C.~Vega, A potential model for
  the study of ices and amorphous water: Tip4p/ice, J. Chem. Phys. 122~(23)
  (2005) 234511.
\newblock \href {http://dx.doi.org/10.1063/1.1931662}
  {\path{doi:10.1063/1.1931662}}.

\bibitem{TavagnaccoPCCP2018}
L.~Tavagnacco, E.~Zaccarelli, E.~Chiessi, On the molecular origin of the
  cooperative coil-to-globule transition of poly(n-isopropylacrylamide) in
  water, Physical Chemistry Chemical Physics 20 (2018) 9997--10010.
\newblock \href {http://dx.doi.org/10.1039/C8CP00537K}
  {\path{doi:10.1039/C8CP00537K}}.

\bibitem{TavagnaccoJPCL2019}
L.~Tavagnacco, E.~Chiessi, M.~Zanatta, A.~Orecchini, E.~Zaccarelli,
  Water--polymer coupling induces a dynamical transition in microgels, The
  Journal of Physical Chemistry Letters 10~(4) (2019) 870--876.
\newblock \href {http://dx.doi.org/10.1021/acs.jpclett.9b00190}
  {\path{doi:10.1021/acs.jpclett.9b00190}}.

\bibitem{TavagnaccoPRR2021}
L.~Tavagnacco, M.~Zanatta, E.~Buratti, B.~Rosi, B.~Frick, F.~Natali,
  J.~Ollivier, E.~Chiessi, M.~Bertoldo, E.~Zaccarelli, A.~Orecchini,
  Proteinlike dynamical transition of hydrated polymer chains, Physical Review
  Research 3 (2021) 013191.
\newblock \href {http://dx.doi.org/10.1103/PhysRevResearch.3.013191}
  {\path{doi:10.1103/PhysRevResearch.3.013191}}.

\bibitem{AbrahamSX2015}
M.~J. Abraham, T.~Murtola, R.~Schulz, S.~Páll, J.~C. Smith, B.~Hess,
  E.~Lindahl, Gromacs: High performance molecular simulations through
  multi-level parallelism from laptops to supercomputers, SoftwareX 1-2 (2015)
  19 -- 25.
\newblock \href {http://dx.doi.org/https://doi.org/10.1016/j.softx.2015.06.001}
  {\path{doi:https://doi.org/10.1016/j.softx.2015.06.001}}.

\bibitem{HumphreyJMG1996}
W.~Humphrey, A.~Dalke, K.~Schulten, Vmd: Visual molecular dynamics, Journal of
  Molecular Graphics 14~(1) (1996) 33 -- 38.
\newblock \href
  {http://dx.doi.org/https://doi.org/10.1016/0263-7855(96)00018-5}
  {\path{doi:https://doi.org/10.1016/0263-7855(96)00018-5}}.

\bibitem{BondiJPC1964}
A.~Bondi, Van der waals volumes and radii, Journal of Physical Chemistry 68
  (1964) 441--451.
\newblock \href {http://dx.doi.org/10.1021/j100785a001}
  {\path{doi:10.1021/j100785a001}}.

\bibitem{EisenhaberCC1995}
F.~Eisenhaber, P.~Lijnzaad, P.~Agros, C.~Sander, M.~Scahrf, The double cubic
  lattice method: Efficient approaches to numerical integration of surface area
  and volume and to dot surface contouring of molecular assemblies, Journal of
  Computational Chemistry 16~(3) (1995) 273--284.
\newblock \href {http://dx.doi.org/10.1002/jcc.540160303}
  {\path{doi:10.1002/jcc.540160303}}.

\bibitem{GallinaJCP2010}
M.~E. Gallina, L.~Comez, A.~Morresi, M.~Paolantoni, S.~Perticaroli, P.~Sassi,
  D.~Fioretto, Rotational dynamics of trehalose in aqueous solutions studied by
  depolarized light scattering, The Journal of Chemical Physics 132~(21) (2010)
  214508.
\newblock \href {http://dx.doi.org/10.1063/1.3430555}
  {\path{doi:10.1063/1.3430555}}.

\bibitem{PrivalovJMB1999}
J.~Taylor, N.~Greenfield, B.~Wu, P.~Privalov, A calorimetric study of the
  folding-unfolding of an alpha-helix with covalently closed n and c-terminal
  loops, Journal of Molecular Biology 291 (1999) 965--976.
\newblock \href {http://dx.doi.org/10.1006/jmbi.1999.3025}
  {\path{doi:10.1006/jmbi.1999.3025}}.

\bibitem{PersikovPS2004}
A.~V. Persikov, Y.~Xu, B.~Brodsky, Equilibrium thermal transitions of collagen
  model peptides, Protein Science 13 (2004) 893--902.
\newblock \href {http://dx.doi.org/10.1110/ps.03501704}
  {\path{doi:10.1110/ps.03501704}}.

\bibitem{FucinosPLOS2014}
C.~Fuci{\~n}os, P.~Fuci{\~n}os, M.~M\'iguel, I.~Katime, L.~M. Pastrana, M.~L.
  R\'ua, Temperature- and ph-sensitive nanohydrogels of
  poly($n$-isopropylacrylamide) for food packaging applications: modelling the
  swelling-collapse behaviour, PLOS ONE 9 (2014) e87190.
\newblock \href {http://dx.doi.org/10.1371/journal.pone.0087190}
  {\path{doi:10.1371/journal.pone.0087190}}.

\bibitem{FutscherSCIREP2017}
M.~H. Futscher, M.~Philipp, P.~M\"uller-Buschbaum, A.~Schulte, The role of
  backbone hydration of poly(n-isopropyl acrylamide across the volume phase
  transition compared to its monomer, Scientific Reports 7 (2017) 17012.
\newblock \href {http://dx.doi.org/10.1038/s41598-017-17272-7}
  {\path{doi:10.1038/s41598-017-17272-7}}.

\bibitem{LiuCHEMCOMM2018}
Z.~Liu, Q.~Gao, J.~Chen, J.~Deng, K.~Lin, X.~Xing, Negative thermal expansion
  in molecular materials, Chemical Communications 54 (2018) 5164--5176.
\newblock \href {http://dx.doi.org/10.1039/C8CC01153B}
  {\path{doi:10.1039/C8CC01153B}}.

\bibitem{LopezSM2017}
C.~G. Lopez, W.~Richtering, Does flory--rehner theory quantitatively describe
  the swelling of thermoresponsive microgels?, Soft Matter 13~(44) (2017)
  8271--8280.
\newblock \href {http://dx.doi.org/10.1039/c7sm01274h}
  {\path{doi:10.1039/c7sm01274h}}.

\bibitem{BischofbergerSCIREP2015}
I.~Bischofberger, V.~Trappe, New aspects in the phase behaviour of
  poly-n-isopropyl acrylamide: systematic temperature dependent shrinking of
  pnipam assemblies well beyond the lcst, Scientific Reports 5~(1) (2015)
  15520.
\newblock \href {http://dx.doi.org/10.1038/srep15520}
  {\path{doi:10.1038/srep15520}}.

\bibitem{NinarelloMACRO2019}
A.~Ninarello, J.~J. Crassous, D.~Paloli, F.~Camerin, N.~Gnan, L.~Rovigatti,
  P.~Schurtenberger, E.~Zaccarelli, Modeling microgels with a controlled
  structure across the volume phase transition, Macromolecules 52~(20) (2019)
  7584--7592.
\newblock \href {http://dx.doi.org/10.1021/acs.macromol.9b01122}
  {\path{doi:10.1021/acs.macromol.9b01122}}.

\bibitem{WoodwardEPJ2000}
N.~C. Woodward, B.~Z. Chowdhry, S.~A. Leharne, M.~J. Snowden, The interaction
  of sodium dodecyl sulphate with colloidal microgel particles, European
  Polymer Journal 36~(7) (2000) 1355 -- 1364.
\newblock \href {http://dx.doi.org/10.1016/S0014-3057(99)00207-4}
  {\path{doi:10.1016/S0014-3057(99)00207-4}}.

\bibitem{BischofbergerSCIREP2014}
I.~Bischofberger, D.~C.~E. Calzolari, P.~D.~L. Rios, I.~Jelezarov, V.~Trappe,
  Hydrophobic hydration of poly-n-isopropyl acrylamide: a matter of the mean
  energetic state of water, Scientific Reports 4 (2014) 4377.

\bibitem{SchildJPC1990}
H.~G. Schild, D.~A. Tirrell, Microcalorimetric detection of lower critical
  solution temperatures in aqueous polymer solutions, Journal of Physical
  Chemistry 94~(10) (1990) 4352--4356.
\newblock \href {http://dx.doi.org/10.1021/j100373a088}
  {\path{doi:10.1021/j100373a088}}.

\bibitem{KujaraMACRO2001}
P.~Kujawa, F.~M. Winnik, Volumetric studies of aqueous polymer solutions using
  pressure perturbation calorimetry: A new look at the temperature-induced
  phase transition of poly(n-isopropylacrylamide) in water and d2o,
  Macromolecules 34~(12) (2001) 4130--4135.
\newblock \href {http://dx.doi.org/10.1021/ma002082h}
  {\path{doi:10.1021/ma002082h}}.

\bibitem{FujishigeJPC1989}
S.~Fujishige, K.~Kubota, I.~Ando, Phase transition of aqueous solutions of
  poly(n-isopropylacrylamide) and poly(n-isopropylmethacrylamide), The Journal
  of Physical Chemistry 93~(8) (1989) 3311--3313.
\newblock \href {http://dx.doi.org/10.1021/j100345a085}
  {\path{doi:10.1021/j100345a085}}.

\bibitem{DingMACRO2005}
Y.~Ding, X.~Ye, G.~Zhang, Microcalorimetric investigation on aggregation and
  dissolution of poly($n$-isopropylacrylamide) chains in water, Macromolecules
  38 (2005) 904--908.
\newblock \href {http://dx.doi.org/10.1021/ma048460q}
  {\path{doi:10.1021/ma048460q}}.

\bibitem{SunSM2013}
S.~Sun, P.~Wu, W.~Zhang, W.~Zhang, X.~Zhu, Effect of structural constraint on
  dynamic self-assembly behavior of pnipam-based nonlinear multihydrophilic
  block copolymers, Soft Matter 9 (2013) 1807--1816.
\newblock \href {http://dx.doi.org/10.1039/C2SM27183D}
  {\path{doi:10.1039/C2SM27183D}}.

\bibitem{JethvaJPCB2017}
P.~N. Jethva, J.~B. Udgaonkar, Modulation of the extent of cooperative
  structural change during protein folding by chemical denaturant, The Journal
  of Physical Chemistry B 121~(35) (2017) 8263--8275.
\newblock \href {http://dx.doi.org/10.1021/acs.jpcb.7b04473}
  {\path{doi:10.1021/acs.jpcb.7b04473}}.

\bibitem{JethvaBIOCHEM2018}
P.~N. Jethva, J.~B. Udgaonkar, The osmolyte tmao modulates protein folding
  cooperativity by altering global protein stability, Biochemistry 57~(40)
  (2018) 5851--5863.
\newblock \href {http://dx.doi.org/10.1021/acs.biochem.8b00698}
  {\path{doi:10.1021/acs.biochem.8b00698}}.

\bibitem{MaedaLANG2000}
Y.~Maeda, T.~Higuchi, I.~Ikeda, Change in hydration state during the
  coil-globule transition of aqueous solutions of poly($n$-isopropylacrylamide
  as evidenced by ftir spectroscopy, Langmuir 16 (2000) 7503--7509.
\newblock \href {http://dx.doi.org/10.1021/la0001575}
  {\path{doi:10.1021/la0001575}}.

\bibitem{AhmedJPCB2009}
Z.~Ahmed, E.~A. Gooding, K.~V. Pimenov, L.~Wang, S.~A. Asher, Uv resonance
  raman determination of molecular mechanism of poly($n$-isopropylacrylamide)
  volume phase transition, Journal of Physical Chemistry B 113 (2009)
  4248--4256.
\newblock \href {http://dx.doi.org/10.1021/jp810685g}
  {\path{doi:10.1021/jp810685g}}.

\bibitem{MaitiJACS2004}
N.~C. Maiti, M.~M. Apetri, M.~G. Zagorski, P.~R. Carey, V.~E. Anderson, Raman
  spectroscopic characterization of secondary structure in natively unfolded
  proteins: $\alpha$-synuclein, Journal of the American Chemical Society
  126~(8) (2004) 2399--2408.
\newblock \href {http://dx.doi.org/10.1021/ja0356176}
  {\path{doi:10.1021/ja0356176}}.

\bibitem{SunMACRO2008}
B.~Sun, Y.~Lin, P.~Wu, H.~W. Siesler, A ftir and 2d-ir spectroscopic study on
  the microdynamics phase separation mechanism of the
  poly(n-isopropylacrylamide) aqueous solution, Macromolecules 41~(4) (2008)
  1512--1520.
\newblock \href {http://dx.doi.org/10.1021/ma702062h}
  {\path{doi:10.1021/ma702062h}}.

\bibitem{PeltonJCIS2010}
R.~Pelton, Poly(n-isopropylacrylamide)(pnipam) is never hydrophobic, Journal of
  Colloid and Interface Science 348~(2) (2010) 673--674.
\newblock \href {http://dx.doi.org/10.1016/j.jcis.2010.05.034}
  {\path{doi:10.1016/j.jcis.2010.05.034}}.

\bibitem{CourtenayBIOCHEM2000}
E.~S. Courtenay, M.~W. Capp, C.~F. Anderson, M.~T. Record~Jr., Vapor pressure
  osmometry studies of osmolyte-protein interactions: Implications for the
  action of osmoprotectants in vivo and for the interpretation of ``osmotic
  stress'' experiments in vitro, Biochemistry 39~(15) (2000) 4455--4471.
\newblock \href {http://dx.doi.org/10.1021/bi992887l}
  {\path{doi:10.1021/bi992887l}}.

\bibitem{MillerJPCB2000}
D.~P. Miller, J.~J. de~Pablo, Calorimetric solution properties of simple
  saccharides and their significance for the stabilization of biological
  structure and function, The Journal of Physical Chemistry B 104~(37) (2000)
  8876--8883.
\newblock \href {http://dx.doi.org/10.1021/jp000807d}
  {\path{doi:10.1021/jp000807d}}.

\bibitem{LupiJPCB2012}
L.~Lupi, L.~Comez, M.~Paolantoni, S.~Perticaroli, P.~Sassi, A.~Morresi, B.~M.
  Ladanyi, D.~Fioretto, Hydration and aggregation in mono- and disaccharide
  aqueous solutions by gigahertz-to-terahertz light scattering and molecular
  dynamics simulations, The Journal of Physical Chemistry B 116~(51) (2012)
  14760--14767.
\newblock \href {http://dx.doi.org/10.1021/jp3079869}
  {\path{doi:10.1021/jp3079869}}.

\bibitem{FiorettoFOOD2013}
D.~Fioretto, L.~Comez, S.~Corezzi, M.~Paolantoni, P.~Sassi, A.~Morresi, Solvent
  sharing models for non-interacting solute molecules: The case of glucose and
  trehalose water solutions, Food Biophysics 8~(3) (2013) 177--182.
\newblock \href {http://dx.doi.org/10.1007/s11483-013-9306-3}
  {\path{doi:10.1007/s11483-013-9306-3}}.

\bibitem{ComezJPCL2013}
L.~Comez, L.~Lupi, A.~Morresi, M.~Paolantoni, P.~Sassi, D.~Fioretto, More is
  different: Experimental results on the effect of biomolecules on the dynamics
  of hydration water, The Journal of Physical Chemistry Letters 4~(7) (2013)
  1188--1192.
\newblock \href {http://dx.doi.org/10.1021/jz400360v}
  {\path{doi:10.1021/jz400360v}}.

\bibitem{TavagnaccoJML2020}
L.~Tavagnacco, E.~Zaccarelli, E.~Chiessi, Molecular description of the coil-to
  globule transition of poly(n-isopropylacrylamide) in water/ethanol mixture at
  low alcohol concentration, Journal of Molecular Liquids 297 (2020) 111928.
\newblock \href {http://dx.doi.org/10.1016/j.molliq.2019.111928}
  {\path{doi:10.1016/j.molliq.2019.111928}}.

\bibitem{CroweCRYO1990}
J.~H. Crowe, J.~F. Carpenter, L.~M. Crowe, T.~J. Anchordoguy, Are freezing and
  dehydration similar stress vectors? a comparison of modes of interaction of
  stabilizing solutes with biomolecules, Cryobiology 27~(3) (1990) 219--231.
\newblock \href {http://dx.doi.org/10.1016/0011-2240(90)90023-W}
  {\path{doi:10.1016/0011-2240(90)90023-W}}.

\end{thebibliography}

\begin{thebibliography}{}

\bibitem{BernePecora} Bruce J. Berne and Robert Pecora, \textit{Dynamic Light Scattering: With
Applications to Chemistry, Biology, and Physics}, Dover
Publications, New York (2000)

\bibitem{1} M. Philipp, U. Muller, R. Aleksandrova, R. Sanctuary,
P. Muller-Buschbaum, J. K. Kruger, On the elastic nature of the
demixing transition of aqueous PNIPAM solutions, \textit{Soft
Matter} \textbf{8}, 11387-11395 (2012)

\bibitem{2} L.Korson, W.Drost-Hansen, F. J. Millero, Viscosity of
water at various temperatures, \textit{J. Phys. Chem.} \textbf{73},
34-39 (1969)

\bibitem{FloryJACS1966} P. J. Flory, J. E. Mark, A. Abe, Random-Coil Configurations of
Vinyl Polymer Chains. The Influence of Stereoregularity on the
Average Dimensions, \textit{J. Am. Chem. Soc.} \textbf{88}, 639-650
(1966)

\bibitem{HockneyMCP1970} R. W. Hockney, The potential calculation and some
applications, \textit{Methods Comput. Phys.} \textbf{9}, 136 (1970)

\bibitem{HessJCC1997} B. Hess, H. Bekker, H. J. C. Berendsen, J. G. E. M. Fraaije,
LINCS: A linear constraint solver for molecular simulations, \textit{J. Comput. Chem.} \textbf{18}, 1463–1472
(1997)

\bibitem{BussiJCP2007} G. Bussi, D. Donadio, M. Parrinello, Canonical sampling through velocity
rescaling, \textit{J. Chem. Phys.} \textbf{126}, 014101 (2007)

\bibitem{ParrinelloJAP1981} M. Parrinello, A. Rahman, Polymorphic transitions in single crystals: A new molecular dynamics
method, \textit{J. Appl. Phys.} \textbf{52}, 7182--7190 (1981)

\bibitem{NoseMP1983} S. Nos{\'e}, M. L. Klein, \textit{Mol. Phys.}
\textbf{50}, 1055--1076 (1983)

\bibitem{EssmannJCP1995} U. Essmann, L. Perera, M. L. Berkowitz, T. Darden, H. Lee, L. G. Pedersen, A smooth particle mesh Ewald
method, \textit{J. Chem. Phys.} \textbf{103}, 8577--8593 (1995)



\end{thebibliography}
\end{document}